# FEcMD: A multi-physics and multi-scale computational program for dynamic coupling molecular dynamics simulations with transient electric field and heat conduction in metal nanostructures

Bing Xiao [a*], Nan Li [b*], Wenqian Kong [a], Rui Chu [a], Hongyu Zhang [a], Guodong Meng [a], Kai Wu [a], Yonghong Cheng [a*]

[a] School of Electrical Engineering, State Key Laboratory of Electrical Insulation and Power Equipment, Xi'an Jiaotong University, Xi'an 710049, China.

[b] New Energy Technology Department, Xi'an Thermal Power Research Institute, Shaanxi, Xi'an, 710032, China

[*] Corresponding author

E-mail: bingxiao84@xjtu.edu.cn, linan@tpri.com.cn, cyh@mail.xjtu.edu.cn.

**Abstract**

Field emission coupled with molecular dynamics simulation (FEcMD) software package is a computational tool for studying atomic structure evolution, structural deformation, phase transitions, recrystallization as well as electron emission characteristics of micro- and nano-protrusions and nanowires consisting of elemental metals or multi-component alloys by means of multi-physics and multi-scale methodology. Current implementations of molecular dynamics simulation coupled with multi-scale electrodynamics (ED) and heat conduction (HC) in FEcMD program are advanced mainly in the two aspects as follows. In electrodynamics, the FEcMD program incorporates the space charge interactions (space charge potential and exchange-correlation effects) in the self-consistent solved Poisson-Schrödinger equation with Wentzel-Kramers-Brillouin-Jeffreys (WKBJ) approximation to evaluate the field emission current density and the related resistive heating process more reliably for nanowires or nano-protrusions especially for nano-gaps between two metal electrodes. Meanwhile, the two-temperature heat conduction model is implemented in electrodynamics coupled with molecular dynamics simulations (ED-MD), providing more dedicated descriptions for the hierarchical electron-phonon two-channel heat conduction mechanism and the temperature evolutions of electron and phonon subsystems under the radiofrequency (RF) or pulse electric fields. Benchmark tests are performed for some key implementations in FEcMD software to validate the numerical results, and also to demonstrate the use of program to study the atomic structure evolution of metal nano-structures under electric field and heating processes.

**Program summary**

*Program* Title: FEcMD

*CPC Library* link *to program files:*

*Developer's* repository *link:* https://github.com/NanLi-xjtu/FEcMD.git

*Licensing* provisions: MIT

*Programming* language: C/C++

*Nature of problem:* Modelling the atomic structure evolutions such as structural deformation, crystalline-to-liquid phase transitions, recrystallization and thermal or field evaporation of metallic nanowires and nano-protrusions under the applied E-field stress and strong transient resistive heating embraces a novel hybrid electrodynamics coupled molecular dynamics simulation scheme and workflow to synergistically update both atomic positions and physical fields (temperature and E-field) within multi-physics and multi-scale simulation methodology.

*Solution method:* The 3-D Poisson equation for electrostatics and two-temperature model for heat conduction are numerically solved on finite element or difference grid for the continuum domain, and which are dynamically coupled with atomic structure evolution obtained from classic molecular dynamics simulations for the atomistic domain, enabling the simultaneous study of transient multi-physical field and atomic structure evolutions of metallic micro- and nano-tips under both static and time-dependent electric fields.

**1. Introduction**

Investigating and understanding structural changes of materials such as elastic/plastic deformation, polymorphic phase transition, melting, recrystallization and thermal/field evaporation under the strong transient resistive heating and the high electrical field stress conditions are now the cutting-edge field to many disciplines in materials science, electrical engineering and condensed matter physics. Unlike the conventional thermally activated structural changes of materials, the atomic structure evolution under the applied E-field is expected to be strongly regulated not only by electric field stresses known as Maxwell stress, but also by the presence of resistive heating process related to field electron emission mechanism especially at high field strength [[1[6]. In fact, optimizing field electron emission performance of metal nanotip plays a crucial role in various applications in modern era, such as electron microscopies [[1, [8], X-ray tubes [[9], high power microwave sources [[10], flat-panel displayers [[11, [12], MEMS systems [[13], electron lithography [[14, [15], fusion reactors [[16] and particle accelerators [[17, [18], etc. Otherwise, utilizing the resistive transient heating under the short electric pulse to realize the unusual phase transformation and to stabilize the metastable phases (i.e., metallic glasses) have been also demonstrated in literature [[1,[2].

Experimentally, it has been well documented that under the ultra-high electric field, field electron emission could trigger thermal runaway event at the surface of electrode, eventually leading to the development of a vacuum arc in the spacing between two electrodes and resulting in the electric breakdown and other irreversible structural damages to the field emission devices or power equipment. In the past several decades, extensive experimental studies have been carried out to reveal the dynamic evolution of microstructure or atomic structure of field electron emitters in terms of micro- or nano-protrusion on the metal electrode surfaces under high E-field strength [[19-[21]. Specifically, transmission electron microscopy (TEM) and scanning electron microscopy (SEM) equipped with high-voltage probes and optical spectrometer could provide the ability to observe the deformation and structural phase transition of the micro- or nano-tips in-situ before and after the vacuum electric breakdown [[22-[25]. However, neither TEM nor SEM could track the microstructure changes of nanotips or nano-protrusions during the electric breakdown or at the pre-breakdown moment because of the intense electromagnetic interferences. Otherwise, the vacuum breakdown usually occurs on an extremely short time scale, it is almost impossible to dynamically track a transient phenomenon such as the atomic-scale structure damages during the electric breakdown process experimentally. Although the structural damages at the atomistic scale and the associated thermal runaway process are long believed to initiate vacuum electric breakdown, detailed atomic structure changes under the high E-field and strong resistive heating are still not fully understood.

Computer simulations based on either molecular dynamics simulations or finite element electromagnetic hydrodynamic calculations are considered as the state-of-the-art methodologies to investigate the dynamic coupling of structural evolution of materials with the Joule heating and electric stress at different stages of electric breakdown [[26, [27]. In a typical finite element hydrodynamic simulation of micro- or nano-protrusions under the high electric field the field electron emission process is usually described by the classic Fowler-Nordheim equation or semi-classic quantum mechanical model [28] after knowing the local electric field distribution by solving either the Poisson equation or Laplace equation. In addition, the Joule and Nottingham heating mechanisms in the interior of electron emitters can be investigated through the heat balance equation involving heat conduction, convection and radiation terms. Notably, hydrodynamic simulations allow the detailed inspection of plastic deformation, melting and evaporation of materials using the laws of conservation of mass, momentum and energy [29[32]. Nevertheless, conventional electromagnetic hydrodynamic simulations and other similar multi-physics schemes using the finite element techniques obviously lack the capability to thoroughly address the multi-scale aspect of structural deformation, phase transition and thermal evaporation of micro- or nano-tips under the influences of transient strong heating and high electric field stress. Obviously, molecular dynamics simulations based on classic interatomic potentials or machine learning potentials are highly efficient to model the atomic structures of materials at various temperatures, pressures and stress conditions [33[36]. Employing the classic MD simulations with the applied electric field remains very difficult for most dielectric materials due to high complexity in the polarization processes and also the lack of polarizable interatomic potentials or force fields [37[39]. Meanwhile, the above issues are largely absent for metals and metallic systems because the external E-field is fully shielded by the conduction electrons on the nanotip or nanowire surface. Therefore, an earlier attempt at addressing the atomic structure evolutions of metal nanowires, nanotips and nano-protrusions under the high E-field was the direct coupling of classic molecular dynamics simulation for discretized particle dynamics with the COMSOL-based multiphysics model of electrostatics [40]. The method could capture the essential physics of thermal runaway process of metal nanotips under high Maxwell stress; however, the field electron emission process and the related heating mechanisms (Joule and Nottingham heats) were not considered in the methodology. Later on, Zhang and coworkers followed a similar approach but using an in-house finite difference method (FDM) to calculate the field electron emission current density as well as the Nottingham heating for the metal nanotips [41]. Otherwise, electric field forces acting on the surface atoms of metal nanotips were constrained to show the axial symmetry in Ref [41] for simplifying numerical calculations, leading to interesting axial symmetric plastic deformation behavior.

    Recent breakthrough in the development of advanced multiscale-multiphysics computational method to modeling the dynamic coupling of E-field with the atomic structure evolution was the devising the hybrid molecular dynamics simulation known as electrodynamics coupled with molecular dynamics (ED-MD) by Djurabekova and coworkers [42-[44]. This revolutionary methodology featured a

multiscale self-adaptive finite element description of spatial electric field evolution that is dynamically coupled with the atomic structure of metal nanotips or nano-protrusions with the simultaneous incorporating of field electron emission current, resistive Joule heating, Nottingham heat and space charge screening of local E-field in a unified workflow, providing the ability to quantitatively understand the correlation between electric pre-breakdown and microstructure damages caused by the intense field emission process [[45]. Using ED-MD-PIC method, Veske and coworkers for the first time characterized the structural evolution of a conic Cu nanotips in multiple successive thermal runaway events under the high E-field [[45]. Previously, we have also applied both ED-MD and ED-MD-PIC techniques to understand the structural damages and pre-breakdown electric properties of metal nanotips [[46-[49]. Nevertheless, the current implementations of ED-MD or ED-MD-PIC methodology showed some limitations regarding the space charge screening effects, the dual channel heat conduction mechanism and the supporting of machine learning potentials, requiring further integration and extension of the algorithms to properly address particularly following aspects.

First, Uimanov and coworkers [[54] demonstrated the electrodynamics coupled with the two-temperature heat conduction finite element simulation to investigate the electrode temperature evolution and the RF vacuum electric breakdown characteristics of micro-protrusions. It has been revealed that phonon and electron thermal conductions should be treated separately for the pulsed heating process at nanosecond range mainly because the thermal energy exchange between phonons and electrons is regulated by the electron-phonon coupling constant. To be more specifically, the electron temperature and phonon temperature must be obtained from the two coupled non-linear thermal conduction equations. For metallic nano-tips or nano-protrusions, the electric pre-breakdown could occur at the time scale shorter than nanosecond or even down to several picoseconds [[46-[49, [55, [56], probably resulting in the highly thermal non-equilibrium conditions for electron and phonon subsystems. Therefore, the realization of two-temperature model in ED-MD-PIC methodology certainly could be vital to provide more accurate physical insights regarding the thermal processes in the metal nanotips under RF or pulsed voltage.

Second, for intense electron emission process in nano-gaps, the space charge density can be large. As a result, the de Broglie wavelength and electrostatic potential energy of electrons are comparable to gap size and space charge energy, respectively. Under such conditions, electrons must be treated as the quantum particles, implying that the space charge potential and exchange-correlation potential of space charge (electrons) cannot be ignored when computing the space charge effects on the local E-field distribution and electron emission process. The quantum effects of space charge on the field emission current have been theoretically addressed by Ang and coworkers [[57, [58] using a mean-field quantum mechanical model based on Kohn-Sham density functional theory (DFT). Specifically, the quantum Child-Langmuir law was derived for field emission current density in the cases of 1D and 2D nano-gaps [[57]. The existing ED-MD-PIC methodology [[26, [27, [46, [47, [60-[63] does not consider the space charge fields (space charge potential and exchange-correlation effects) among emitted electrons in calculating the electron transmission coefficient using WKBJ approximation in Kemble formula. Although space charge fields are negligible for micro-size gaps applying to typical filed emitters, the inclusion of space charge fields in the electron emission model is anticipated to significantly improve the accuracy and reliability of ED-MD-PIC simulations for electrodynamics and structural evolution when the spacing the between cathode and anode is in the submicron- or nano-range.

In this paper, we presented recent key efforts to address the dynamic coupling of atomic structures of metallic nanotips or nano-protrusions with the transient local electric fields and heating processes under the unified workflow in the developed computational tool known as field emission coupled with molecular dynamics (FEcMD) code. FEcMD code was designated to provide a self-contained and highly integrated platform for realizing state-of-the-art multiscale-multiphysics hybrid molecular dynamics simulations for metallic nanotips and nano-protrusions (or nanowires). Comparing with some existing programs and codes in the relevant fields, main advancements in current algorithms of FEcMD program included the dynamic mapping phonon and electron temperature profiles to atomic structure from the two-temperature model in heat conduction simulation, solving the coupled Poisson-Schrödinger equation for semi-classical electrons with the inclusion of space charge fields (space charge Hartree potential and exchange-correlation potential) using density functional theory (DFT), and the interface to support machine learning potentials. In general, FEcMD program could be considered as a highly versatile computational tool to study the mechanical properties, phase transition and field electron emission characteristics of metal nanotips or nanowires under the local electric fields and heating process from a hierarchal multiscale-multiphysics methodology.

The present paper is organized as follows: The overall workflow of FEcMD program is discussed in section 2. In section 3, the theoretical methods and backgrounds for different computational modules are presented. The implementation details of molecular dynamics simulation, two-temperature model for heat conduction, and the space charge quantum effects are addressed in section 4. In the same section, we demonstrate some typical applications of FEcMD program to model atomic structure of micro- or nano-protrusions under E-field and heating process. Regarding some recent applications of FEcMD for investigating the atomic structure evolution of metal nanotips under electric field and heating processes, we refer to Ref [[64, [65].

## 2. Computational modules in FEcMD code

The workflow of FEcMD program is fully packed into four computational modules including molecular dynamics (MD), atomic forces (AF), electrodynamics (ED), and heat conduction (HC), as illustrated in Figure 1. The additional finite-element mesh (FEM) generation module mainly serves for HC and ED modules to solve the heat balance equation for heat conduction and Poisson equation for electrodynamics. The FEM module also plays the key role in FEcMD software to dynamically couple the atomic structure of nanotip obtained from MD module with HC and ED modules that determine the temperature and E-field distributions, supporting the multi-scale and multi-physics simulations for nano-emitters using ED-MD-PIC methodology. To perform the full ED-MD-PIC simulation, all four major computational modules presented in Figure 1 are employed and executed consecutively, i.e., MESH → ED+HC → AF+MD → MESH → ED-HC and etc. To be more specific, the MESH module employs the input atomic structures of nanotips to generate the muti-scale finite element meshes in the vacuum region, the surface and the interior of nanotip, then ED and HC modules utilize the finite element meshes and the proper initial and boundary conditions to obtain the local electric fields and temperature profile. Atomic charges of surface atoms are further calculated from the local electric field strength from Gauss law, and which are used to compute the additional electric forces acting on those atoms besides other intrinsic interatomic interactions that are described by classic force fields or machine learning potentials. Meanwhile, the local temperature distribution obtained from HC module in the interior of nanotip is directly mapped onto the atoms through the velocity scaling method. Perturbations in both atomic forces and velocities are seamlessly incorporated into MD module for updating the atomic positions. The whole simulation cycle repeats until reaching the user designated maximum number of MD steps.

Otherwise, FEcMD software supports the use of MD + AF modules, HC + MESH modules, ED + MESH modules as stand-alone computational tools to perform the molecular dynamics simulations, heat conduction and electrodynamics, respectively. Besides the full capability of ED-MD-PIC simulation, no further modification of the FEcMD software for the use of aforementioned standalone features is required from the users, and only some minor changes in the input file are needed. Notably, the stand-alone finite element simulations (HC/ED + MESH modules) in FEcMD program only adopt the atomic positions of nanotips for the initiation of finite element mesh, while the subsequent numerical calculations of electric fields and temperature profile are conducted on the fixed geometry of the nanotip without dynamic updating of mesh points as it is done in ED-MD simulation. Meanwhile, performing the stand-alone molecular dynamics simulation within MD and AF modules is fully decoupled to the remaining modules (HC, ED and MESH) in FEcMD program. In Figure 2, the detailed workflow of FEcMD software and simulation capabilities of each module are illustrated. .

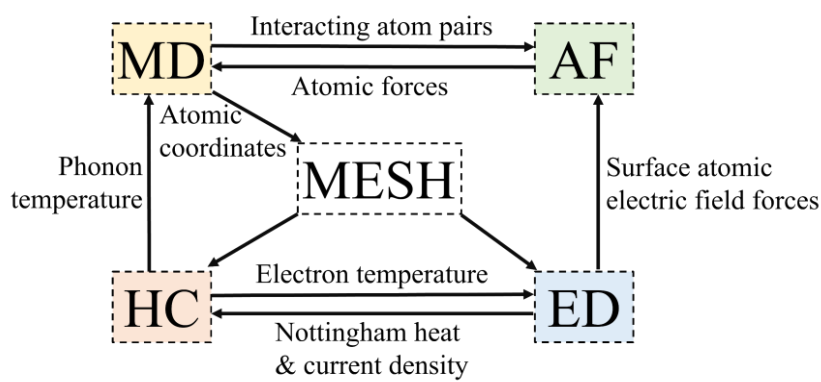

**Figure 1.** Main computational modules and their relationships in FEcMD program.

As shown in Figure. 2, the MD module together with AF module perform the stand-alone molecular dynamics simulations at the finite temperature under various thermodynamic ensembles (NVT, NVT, NPT and non-equilibrium MD) without invoking MESH, ED and HC modules. In the current implementations (*ensemble.c*, *nebrList.c*, *atomposition.c*, *measurements.c*, *eam_alloy.c*, *snap.c* and *mlip.cpp*), the MD module provides a highly versatile interface to support various thermodynamic ensembles (NVE, NVT and NPT) and interatomic potential forms including Lennard-Jones (L-J) pair potential [[66, [67], embedded atomic method (EAM) potential [[68-[70], spectral neighbor analysis potential (SNAP) [[71[72], and moment tensor potential (MTP) [[73, [74]. The AF module is responsible for evaluating both intrinsic inter-atomic forces and external disturbing forces (electric field forces and bare Coulomb force) for bulk or surface atoms. Notably, when performing the standard MD simulation independently, the external atomic forces such as Lorentz and Coulomb forces for charged atoms on the nanotip surface are not calculated. In the AF module, the realization of MLP is supported by the open-source MLIP library (*mlip.cpp*) [[75], while the use of SNAP is interfaced to the *snap.c* subroutine in LAMMPS code [[71] The extraction of surface atoms is realized by performing the atomic coordination number analysis within the density-based spatial clustering of applications with noise algorithm (DBSCAN) [[76[78].

The electrostatic potential and local electric field distributions are obtained from ED module where the Poisson equation is solved numerically in three-dimensional (3D) space on a finite element grid using *VoronoiMesh.cpp* and *PoissonSolver.cpp* subroutines in FEMOCS library [[27]. In ED module, electron emission current density distribution and total emission current density within the space charge quantum effects are calculated using the in-house *EmissionReader.cpp* subroutine which is fully integrated into FEMOCS library [[79]. Space charge density distribution is determined from particle-in-cell (PIC) simulation in ED module, employing the *Pic.cpp* subroutine in FEMOCS library [[80]. The space charge Coulomb potential and exchange-correlation potential are obtained from our new implementations (See section 3 for details) with the integration of open source LIBXC library [[81]. Then the WKBJ approximation and Kemble formula are employed to calculate the electron transmission coefficient and electron emission current density by GETELEC library. It is worth mentioning that the algorithm implemented in GETELEC library is capable of determining the electron emission current density reliably from a smooth transition from thermal emission to cold field electron emission process [[60]. After knowing the electron emission current density distribution on the emitter surface, the Nottingham heat is obtained from GETELEC library. Besides the static E-field, the ED module in FEcMD software also provides the major updates in the algorithm to support ED-MD simulations under the RF E-field and pulsed voltages, using the subroutines *PoissonSolver.cpp* and *EmissionReader.cpp* in FEMOCS library [[83, [84].

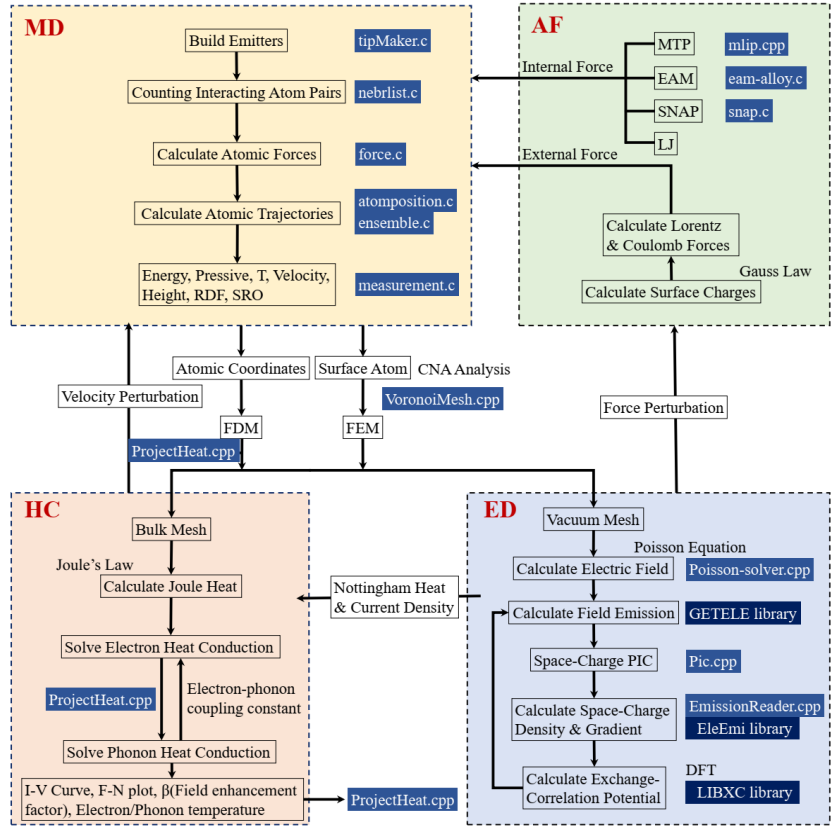

**Figure 2.** Workflow of FEcMD code and the integrated key subroutines and open-source libraries in four main computational modules (MD, AF, ED and HC).

The electric current density and the associated resistive Joule heating are evaluated by HC module in FEcMD program through the in-house *ProjectHeat.cpp* subroutine. In the HC module, the local temperature evolutions in the interior of metal nanotips or nano-protrusions are calculated by numerically solving the heat balance equations of both electrons and phonons on the finite difference grid by means of two-temperature model (TTM), as can be seen from Figure. 2. Meanwhile, solving the two-temperature model for heat conduction with finite difference grid is realized by employing the *ProjectHeat.cpp* subroutine in FEMOCS lib. The local temperature distribution is mapped onto atoms in a designated volume element through the velocity rescaling method in MD simulation. Unlike the conventional one-temperature heat conduction mechanism for phonons, the TTM algorithm solves the electron heat balance equation in advance to that of phonon heat conduction, and the electron thermal energy is transferred to phonon system through the electron-phonon coupling mechanism.

Finally, FEcMD program also includes the post data processing subroutines (*measurements.c* and *ProjectHeat.cpp*) for analyzing structural properties and filed emission characteristics of metal nanotips including atomic coordination number (ACN), atomic radial distribution function (RDF), atomic short-range order parameter (SRO), field electron emission current density/distribution, local E-field distribution, electron/phonon temperature distribution, and multiscale finite element mesh data files. The processed data can be further examined and visualized by using other software and programs such as OVITO [[85]], ParaView [[86]], VMD [[87]], and VESTA [[88]].

## 3. Overview of methods in FEcMD modules

### 3.1 Molecular dynamics simulation

The classical molecular dynamics simulations based on either conventional interatomic potentials (Lennard-Jones potentials and EAM potentials) or machine learning potentials (SNAP and MTP) are implemented in ED and AF modules in FEcMD code. For ED-MD simulations, the evolution of modelling structure is determined by the instantaneous forces acting on the constituting particles, following Newton's second law for the equation of motion, as given by Eq. (1). Here, $m_i$ and $r_i$ denote the mass and position of particle $i$ in the atomistic model, and force components including intrinsic interatomic force, the Lorentz force and Coulomb force are represented by $\vec{f}_\mathrm{I}$, $\vec{f}_\mathrm{L}$ and $\vec{f}_\mathrm{C}$, respectively.

$$m_i \frac{\mathrm{d}^2 \vec{r}_i(t)}{\mathrm{d}t^2} = \vec{f}_\mathrm{I}(\vec{r}_i) + \vec{f}_\mathrm{L}(\vec{r}_i) + \vec{f}_\mathrm{C}(\vec{r}_i) \tag{1}$$

The calculation of intrinsic interatomic forces is dependent on the employed atomic potentials, i.e., L-J potential, EAM and machine learning potentials. Meanwhile, the force components such as Lorentz and Coulomb are only applied to those partially ionized atoms at the tip surface with the applied E-field in ED-MD simulations. In the case of the ordinary stand-alone MD simulations in FEcMD code, the Lorentz and Coulomb forces are set to zero.

For evaluating the Lorentz and Coulomb forces in Eq. (1), AF module collaborates with ED module and the *VoronoiMesh.cpp* lib in FEMOCS to calculate the excessive charges carried by the surface atoms utilizes Gauss's law, because those atoms are partially ionized due to the field electron emission process under high local electric field values [[26]]. Then, surface atoms are expected to experience both inter-atomic Coulomb forces [[44]] and Lorentz forces under the E-field in ED-MD simulation. The total charges of partially ionized surface atoms are calculated from Eq. (2), while the corresponding Lorentz and Coulomb forces are obtained from Eq. (3) and (4):

$$q_i = \sum_{f_{MV}} \vec{F}_f \cdot \hat{n}_f A_f ,  \qquad (2)$$

$$\vec{f}_L(\vec{r}_i) = \frac{1}{2} q_i \vec{F}_i , \qquad (3)$$

$$\vec{f}_C(\vec{r}_i) = \frac{1}{4\pi\varepsilon_0} \sum_{j \neq i} \frac{q_i q_j}{r_{ij}} \hat{r}_{ij} \exp(-\xi r_{ij}) , \qquad (4)$$

here, $F$ refers to the local electric field strength at the particle position $r_i$ on the surface of nanotip, and the term " $1/2$ " implies that only one surface of the volume element is fully in contact with the vacuum, and which is exposed to the electric field flux; The term "$q_i$" means the net charge on each ionized atom according to Gauss's law in Eq. (2); "$f_{MV}$" signifies the fact that for each finite element cell containing a surface atom $i$, and the integration is performed over the product of the local electric field at the volume element face center (or the Voronoi cell facets), the unit vector ($\hat{n}_f$) perpendicular to the cell face, and the area of the face ($A_f$); $r_{ij}$ denotes the interatomic distance between atoms $i$ and $j$ in the atomistic model, and $\hat{r}_{ij}$ refers to the unit vector in the direction of $\vec{r}_{ij}$. As can be seen from Eq. (4), the Coulomb interactions among surface charges are screened by the screening parameter $\xi$, and its value is determined by the materials, as noted in Ref [[45].

The AF module also implements a simple algorithm to compute the Maxwell stress for a flat atomic surface, and further mapping the electric forces to surface atoms under the applied E-field. The algorithm is useful to study the initiation and growth of nano-protrusions on a planar polycrystalline metal slab under high E-field at grain boundaries [[92]. For the planar metal electrode, the Maxwell stress and the corresponding electric field forces acting on surface atoms are given by Eqs. (5) and (6).

$$\sigma = \frac{\varepsilon_0}{2} F^2 \qquad (5)$$

$$\vec{f}_{ex} = \frac{\sigma S}{N} \vec{n} \qquad (6)$$

Where, $\sigma$ denotes the Maxwell stress, and $F$ refers to the electric field strength. $S$ and $N$ represent the surface area and total number of surface atoms for planar electrode, and $\vec{n}$ is the surface normal. In the MD simulation, $f_{ex}$ is included together with intrinsic interatomic forces obtained from interatomic potentials to calculate the net atomic force. The selection of surface atoms is done by conducting the atomic coordination number analysis (CNA) using *measurements.c* subroutine in MD module.

Finally, in the typical ED-MD multi-physics simulation, the macroscopic phonon temperature is obtained by solving the heat balance equations in the nanotips on a finite difference mesh. Then, velocities of atoms are regulated according to temperature profile determined from phonon heat balance equation. The rescaling of particle velocities is realized by Eq. (7) [[89].

$$\vec{v}_i = \gamma \vec{v}_i \tag{7a}$$

$$\gamma = \sqrt{\frac{d \cdot \left(1 - \frac{1}{N}\right) \cdot T}{\sum (m_i v_i)^2 \cdot \frac{1}{N}}} \tag{7b}$$

Where, $\gamma$ refers to the rescaling constant, and $N$ denotes the total number of particles in a particular slice. For 3-dimenisonal MD simulation, $d=3$. Other symbols have their usual physical meanings. More details regarding the implementation of TTM model in FEcMD program are addressed in section 4.1.

**3.2 Electrodynamics**

In ED module, the local electrostatic potentials and electric fields are obtained from FEM solver for Poisson equation, as given by Eqs. (8) and (9), respectively.

$$\nabla^2 \varphi = -\frac{e\rho}{\varepsilon_0} \tag{8}$$

$$\vec{E}(r,z) = -\nabla \varphi(r,z) \tag{9}$$

Besides the static E-field, the ED module also supports the pulsed and RF electric fields, as described by Eqs. (10) and (11). The pulsed E-field is defined by the pulse duration ($t_{max}$) and pulse amplitude ($F_0$). On the other hand, the RF electric field is approximated by sine function with the user defined angular frequency ($\omega$) and amplitude ($F_0$).

$$F = \begin{cases} F_0, & 0 < t < t_{max} \\ 0, & \text{others} \end{cases} \tag{10}$$

$$\begin{aligned} F &= F_0 \sin(\omega t) \\ \omega &= 2\pi f \end{aligned} \tag{11}$$

For the field electron emission process, ED module interfaces directly with the GETELEC library to calculate the field emission current density $J$ and Nottingham heat $P_N$, which are expressed by Eqs. (12) and (13) [[60, [99]. The electron transmission coefficient $D(E)$ is obtained by WKBJ approximation and Kemble formula, as shown in Eqs. (14) and (15).

$$J = Z_s k_B T_e \int_{-\infty}^{\infty} D(E) \log\left(1 + \exp\left(-E/k_B T_e\right)\right) dE \tag{12}$$

$$P_N = Z_S \int_{-\infty}^{\infty} \frac{E}{1 + \exp(E/k_B T)} \int_{-\infty}^{E} D(\xi) d\xi dE \tag{13}$$

$$D(E) = \frac{1}{1 + \exp(G(E))} \tag{14}$$

$$G(E) = g \int_{x_1}^{x_2} \sqrt{U(x) - E} dx \tag{15}$$

Where, the $D(E)$ refers to electron transmission coefficient, and $G(E)$ denotes the Gamow exponent. $U(x)$ represents the electron tunneling barrier. $Z_s \approx 1.618 \times 10^{-4}$ A·(eV·nm)$^{-2}$, is the Sommerfeld current constant and $T_e$ is the electron temperature. In ED module, the electron tunneling barrier $U(x)$ is given as Eq. (16). The first three terms in Eq. (16) represent the work function ($\phi_{WF}$), the applied external electric potential ($V(x)$), and image charge potential ($\varphi_{ic}(x)$). The fourth term accounts for the quantum many-body interactions among the emitted electrons (space charges), and which is calculated from the exchange-correlation potential ($\varphi_{xc}(x)$) in the mean field theory such as density functional theory (DFT). This term must be included in quantum field electron emission regime because of its significant effect on the electron tunnelling energy barrier profile [[58]. The last term in Eq. (16) gives the space charge potential ($\varphi_{sc}(x)$), resembling the Hartree potential of electron density in DFT. The space charge potential and exchange-correlation potential could significantly affect the electron tunneling barrier $U(x)$ when the emission current density is large in the nano-gap [[56, [58].

$$U(x) = \phi_{WF} - eV(x) - \varphi_{ic}(x) + \varphi_{xc}(x) + \varphi_{sc}(x) \tag{16}$$

Regarding the image charge potential ($\varphi_{ic}(x)$) in Eq. (16), the classic expression is written as Eq. (17). The classic form of image charge potential has the singularity at $x = 0$, and calculated electron emission energy barrier profile is unrealistic at the very small distance near the interface of emitter and vacuum. Otherwise, Eq. (17) may only be applicable to the planar electrode.

$$\varphi_{ic}^{classic}(x) = -\frac{e^2}{16\pi\varepsilon_0 x} \tag{17}$$

For the non-planar emitter, Ref [[100] suggested that Eq. (18) may be used to calculate the image charge potential by considering the local radius of curvature (RoC) of metal nanotip. However, Eq. (18) does not eliminate the singularity of image charge potential at the interface between electrode and vacuum ($x = 0$).

$$\varphi_{ic}^{classic}(x) = -\frac{Q}{16\pi\varepsilon_0 x(1+x/2R)} \tag{18}$$

To eliminate the unphysical singularity in image charge potential at the cathode surface, the more elaborated method must be employed. One such method is the use of free electron Thomas-Fermi approximation (TFA) with the random phase approximation (RPA) to calculate the screening effects of image charge potential [[101]. In this method, the image charge potential is obtained from Eq. (19). Here, Green function of a longitudinal self-consistent electric field $D_{vac}(p;x,x')$ describes the screened Coulomb interaction between the charges at points $x$ and $x'$, and $p$ is the wavevector along the $x$ direction (field electron emission direction).

$$\phi_{ic}^{TFA}(x) = -\frac{e^2}{4\pi\varepsilon_0}\int_0^\infty p\mathrm{d}p\left[D_{vac}(p;x,x') + \frac{1}{2p}\right] \tag{19}$$

For calculating the longitudinal self-consistent electric field $D_{vac}(p;x,x')$, Eqs. (20-a)-(20-d) are used, and their derivations can be found in Ref [[101]. In those expressions, $D$ denotes the effective spacing between cathode and anode, and the Thomas-Fermi screening constant $\kappa_0 = 6\pi e^2 n / E_F$ with $n$ the electron density (space charge density), the electron Fermi energy $E_F = \hbar^2(3\pi^2 n)^{2/3}/2m$ and electron rest mass $m$.

$$D_{vac}(p,x,x') = \frac{\gamma_1^S(p;x)}{\gamma_2^S(p)} + \Delta D_{vac}(p;x,x') \tag{20-a}$$

$$\gamma_1^S(p;x) = \sqrt{p^2+\kappa_0^2}\left\{\sqrt{p^2+\kappa_0^2}\begin{bmatrix}\sinh((D-x)p)\cosh((D-x)p)\\+\sinh(xp)\cosh(xp)\end{bmatrix}+p\left[\cosh^2((D-x)p)+\cosh^2(xp)\right]\right\} \quad \text{(20-b)}$$

$$\gamma_2^S(p) = p\sinh(Dp)\left[(2p^2+\kappa_0^2)\sinh(Dp)+2p\sqrt{p^2+\kappa_0^2}\cosh(Dp)\right] \quad \text{(20-c)}$$

$$\Delta D_{vac}(p;x,x') = -\frac{1}{2p}\frac{\{\cosh[(D-2x)p]+\cosh(Dp)\}}{\sinh(Dp)} \quad \text{(20-d)}$$

The exchange-correlation potential of space charges is evaluated using the mean field theory by means of density functional theory (DFT). In DFT, $\varphi_{xc}(x)$ is given by Eq. (21) using local density approximation (LDA) as the typical case. Other advanced exchange-correlation density functionals such as generalized gradient approximation (GGA) and meta-generalized gradient approximation (meta-GGA) have been tested in Ref [107, 108] for field electron emission properties, and no significant difference was revealed in the results of LDA, GGA and meta-GGA forms. Notably, $\varepsilon_{xc}$ represents the exchange-correlation energy per particle for uniform electron gas model in Eq. (21), and its evaluation can be found in literature [109, 81],

$$\varphi_{xc}(x) = \varepsilon_{xc}[n(x)] + \frac{\delta\varepsilon_{xc}[n(x)]}{\delta n(x)} \quad (21)$$

To evaluate the total electron tunneling barrier, as given by Eq. (16), the image charge potential ($\varphi_{ic}(x)$) and the exchange-correlation potential ($\varphi_{xc}(x)$) must be calculated from the space charge density profile ($n(x)$), while the space charge Coulomb potential ($\varphi_{sc}(x)$) is obtained by solving Poisson equation for the space charge distribution. Obviously, those potentials could also affect the instantaneous space charge distribution in the direction of field electron emission. Therefore, the space charge distribution $n(r)$ and the potentials including $\varphi_{ic}(x)$, $\varphi_{xc}(x)$ and $\varphi_{sc}(x)$ must be determined self-consistently. In ED module, the coupled one-dimensional (1-D) Schrödinger-Poisson equation is adopted to calculate the space charge density distribution and various potentials in the field electron emission direction, and which is written as Eq. (22)

$$\left.\begin{array}{l}\dfrac{d^2 q(x)}{dx^2}+\lambda^2\left[\dfrac{\varphi_{sc}(x)-\varphi_{xc}(x)}{eV_g}+\dfrac{x}{D}-\dfrac{4}{9}\dfrac{\mu^2}{q^4}\right]q=0\\[6pt]\dfrac{d^2\varphi_{sc}(x)}{dx^2}=\dfrac{2}{3}\dfrac{eV_g}{D^2}q^2\end{array}\right\}.\qquad(22)$$

Here, $q$ represents the local amplitude of space charge wavefunction, and which is approximated as the plane wave ($\psi(x)$) for semi-classic electrons in terms of Eq. (23), while it is also directly related to the space charge density through Eq. (24). Otherwise, $n_0$ and $\lambda$ denote the electron density scale and normalized gap spacing, respectively, and which are calculated using Eq. (25).

$$\psi(x)=\sqrt{n_0}q(x)\exp[i\theta(x)] \qquad (23)$$

$$n(x)=\psi^*(x)\psi(x)=n_0 q^2(x) \qquad (24)$$

$$\left.\begin{array}{l}n_0=\dfrac{2\varepsilon_0 V_g}{3eD^2}\\[6pt]\lambda=\dfrac{D}{\lambda_0}\end{array}\right\} \qquad (25)$$

Furthermore, in Eqs. (22)-(25), $V_g$ refers to the applied voltage to the effective gap with spacing $D$, and $\varepsilon_0$ the dielectric constant of vacuum; the electron de Broglie wavelength at $V_g$ is given as $\lambda_0=(\hbar^2/2emV_g)^{1/2}$; the phase of the plane wave is given by $\theta(x)$ in Eq. (23), and which has been discussed in Ref [[94]. The normalized electron emission density $\mu=J/J_{CL}$ with $J$ as emission current density obtained from Eq. (12), and $J_{CL}$ represents the classical space-charge limited current density known as Child-Langmuir law in Eq. (26).

$$J_{CL}=\dfrac{4\varepsilon_0}{9}\sqrt{\dfrac{2e}{m}}\dfrac{V_g^{3/2}}{D^2} \qquad (26)$$

The coupled 1D Poisson-Schrödinger equation for space charge density distribution can be numerically solved using the proper boundary conditions which are given by $q(1)=(2\mu/3)^{1/2}$, $q'(1)=0$, $\varphi_{sc}(0)=0$ and $\varphi_{sc}(1)=0$ ( Dirichlet boundary conditions) or $\varphi'_{sc}(0)=s_c$ and $\varphi'_{sc}(1)=s_a$ (Neumann boundary conditions), where the normalized distance $x/\lambda=1$ and $x/\lambda=0$ represent the locations of the virtual anode and the actual cathode in the electron emitting direction for the given μ, respectively. Meanwhile, $s_a$ and $s_c$ denote the electric fields of anode and cathode surfaces [[57]]. Overall, Eqs. (12)-(26) are used in ED module to calculate the field electron emission properties of metal nanotips with and without space charge quantum many-body effects.

Finally, for the non-planar electron emitter, the local electric field employed in Fowler-Nordheim equation to calculate the emission current density must the corrected by considering the local geometry of nanotip. Previously, Kyritsakis and coworkers employed the local radius of curvature (RoC) as the main variable to correct the electrostatic potential for sharp electron emitters down to RoC of 4~5 nm [[60]]. In ED module, solving Poisson equation directly gives the local electrostatic potential values on the finite element mesh with the effects of local curvature on the numerical solutions. Then, the cubic spline interpolation method is used to smoothly evaluate the electrostatic potential along the electron field emission direction on nanotip. To obtain the space charge density distribution in the vacuum region of simulated box, the PIC method is adopted to simulate the collision, and migration of electrons. The incorporation of PIC method in ED module is supported by *Pic.cpp* subroutine in FEMOCS library [[45]]. For more technique details and algorithms regarding PIC simulation in FEMOCS library, we refer to Ref [[45]].

### 3.3 Heat conduction (HC) module

In the case of time-dependent E-field, the electron field emission, and the associated heating processes (Joule and Nottingham heating mechanisms) in micro- or nano-protrusion could last few ns or hundreds of ps, depending on the frequency of RF electromagnetic field or the duration of pulsed voltage (See Eqs. (12) and (13)). The heat conduction mechanism during a transient resistive heating process could be different to that of steady-state heating, especially considering that the phonon-electron energy exchange rate has finite value. In an extreme case where the duration of heating is shorter than that of electron-phonon relaxation time, the dual-channel heat conduction process must be employed to capture the characteristic fast electron heat conduction mechanism for radiofrequency nano-electronic devices [110, 111]. Therefore, the two-temperature employed by Uimanov and coworkers [[54]] in their electrodynamics is implemented in heat conduction (HC) module. In the two-temperature model (TTM), the heat balance equations for phonons and electrons are given by Eqs. (27) and (28), respectively.

$$C_e(T_e)\frac{\partial T_e}{\partial t} = \nabla\left[\kappa_e(T_e,T_p)\nabla T_e\right] - G_{ep}(T_e - T_p) + \rho(T_p)j^2 \tag{27}$$

$$C_p(T_p)\frac{\partial T_p}{\partial t} = \nabla\left[\kappa_p(T_p)\nabla T_p\right] + G_{ep}(T_e - T_p) \tag{28}$$

Here, the subscripts 'e' and 'p' represent electrons and phonons, respectively. $C$, $T$ and $\kappa$ are the volumetric heat capacity, temperature, and thermal conductivity, respectively; $t$ is the heat conduction time; $\rho$ is the resistivity of the emitter; $j$ is current density; $G_{ep}$ is the electron-phonon coupling factor, which determines the energy exchange rate between electrons and phonons. Otherwise, the Nottingham heating of field emission occurs at the tip surface, and this part of the heat flux is given as a boundary condition for heat conduction in Equation (29).

$$-\kappa_e \nabla T_e \big|_{\text{surface}} = q_N \qquad (29)$$

Here, $q_N$ is the Nottingham heat flux, and which is given in Ref [[60]]. In the field emission process, the Nottingham effect can either heat or cool the electron emitter according to the energy released or absorbed by the electrons passing through the potential barrier. The initial temperature of the emitter $T_{e,p}\big|_{t=0} = T_0$, and the temperature of bulk bottom $T_{e,p}\big|_{\text{bottom}} = T_0$.

Some key thermophysical properties of metals must be provided in HC module to solve TTM for electron and phonon subsystems using Eqs. (27) and (28), including specific heat capacities, thermal conductivities and electrical resistivity spanning a wide range of temperatures and phase conditions. In typical ED-MD simulations, electron and phonon temperatures are expected to be substantially higher than the room temperature especially under the high electric field, their dependences on temperature could be simplified. Specifically, the specific heat capacity of phonon $C_p(T_p)$ is taken as a constant, i.e., 3R with R the universal gas constant, obeying the conventional Dulong-Petti law when the temperature is higher than Debye temperature. Meanwhile, the electron specific heat capacity $C_e(T_e)$ is linearly related to electron temperature ($T_e$) by Eq. (30), where $\gamma_e$ is a material dependent parameter, and the values for common metals can be found in literature [[61]].

$$C_e(T_e) = \gamma_e T_e \qquad (30)$$

For thermal conductivities, the electron contribution is estimated by Wiedemann-Franz law using Eq. (31), where the electrical conductivity is denoted as $\sigma$ and the Lorenz number. The most applicable value for Lorenz number is given as $L = 2.44 \times 10^{-8}$ W/($\Omega \cdot$K) for metals. Nevertheless, the precise value of Lorenz number is affected by carrier concentrations, carrier scattering mechanisms and characteristic dimensions of materials, i.e., the value of Cu nanofilm of 40 nm thickness is about $2.00 \times 10^{-8}$ W/($\Omega \cdot$K) [[45]]. For most metals, the total thermal conductivity is dominated by electrons, and the phonon contribution may be designated as a constant or it can be simply set to zero in the TTM simulation.

$$\kappa_e(T_e) = L\sigma(T_e)T_e \quad (31)$$

The electrical resistivities or conductivities for common metals are widely documented in literature for crystalline phase and liquid at some temperatures. Obviously, the tabulated data can be linearly interpolated to estimate the electrical resistivity at any designated temperature for the finite difference grid in the numerical calculations. In the case of solid to liquid phase transition, the electrical resistivity experiences an abrupt change at the melting point, then a similar linear interpolation scheme is used to calculate the apparent electrical resistivity at the phase transition temperature.

**3.4 Boundary conditions**

The multiscale model of metal protrusion is displayed in Figure. 4, and which consists of two different descriptions including the continuum and atomistic domains, as highlighted with small rectangular box containing atomistic model that is embedded in the large continuum region with dimensions $L_x \times L_y \times L_z$. Furthermore, the whole structure of metal protrusion is made of both atomistic and continuum models with the total length as $h_t$. The upper part with the user defined height ($h_a$) is built as atomistic model, while the lower section with length of $h_b$ is treated as coarse-grained domain. The apex of tip is usually characterized by the radius of curvature ($r_0$) and the conical opening angle ($\theta$) for its geometry. The adoption of coarse-grained model for the lower section of metal tip obviously could reduce the computational costs in the ED-MD simulation, compared to that of using full atomistic model for the whole nanotip. This is also a very sound strategy in physics, because the intense field emission process is expected to occur only in the apex region where the local electric fields are sufficiently high for nanotip. The bottom of the tip is further attached to a flat surface representing the bulk surface of metal. As can be seen from Figure 4, the tip structure is placed in a large rectangular box which is periodic in the lateral directions (x and y directions) but aperiodic in the vertical direction (z-direction). In the z-direction, the upper surface of the box represents the boundary of anode, and which is usually assumed to be sufficiently far away from the cathode right below for the conventional electron field emission devices. However, such a requirement has been lifted in the ED module of FEcMD program, because our new implementations of field electron emission model allow the calculation of the field electron emission current density and the associated space charge density distribution from the coupled Poisson-Schrödinger equation for nano-spacing between two electrodes. Both the vertical height and lateral dimensions of the simulation box are set to be significantly larger than that of nano-emitter in the typical field emission simulations, while the use of small height in z-direction is enabled for simulating the field electron emission in nanogaps [[59]. Specifically, in the case of nanogap, the vertical height of the box is determined by total height of nanotip and the applied macroscopic E-field value which determines the spacing between the apex of tip structure and upper anode.

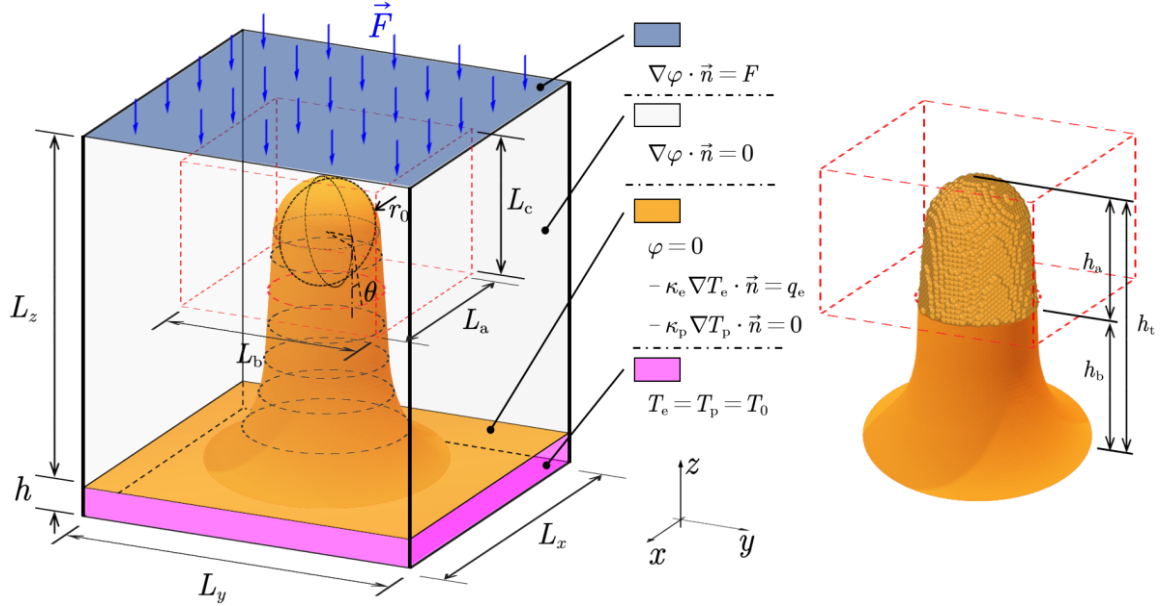

**Figure 4.** The multiscale model of metal nano-protrusion and the multiphysics boundary conditions employed for ED-MD simulations in FEcMD program.

The multiphysics boundary conditions employed for electrodynamics and heat conduction simulations in FEcMD program are illustrated in Figure 4. The mixed Neumann and Dirichlet boundary conditions are used for solving Poisson equation in the continuum description. For example, the top anode is treated as a planar electrode with a fixed E-field value that matches the macroscopic applied E-field (Neumann boundary condition). The surface of a metal nano-tip is equipotential, and the value is set to 0 V (Dirichlet boundary condition). For surfaces of the simulated box in the lateral directions of continuum region, the electric fields in *x* and *y* directions are also set to 0 V (Neumann boundary conditions).

For heat conduction simulation based on finite difference method (FDM), both phonon and electron heat balance equations are solved inside the metal nano-protrusion for both atomistic and continuum domains (the coarse-grained region). The initial temperature of the whole simulation domain (continuum and atomistic descriptions) is set to the user specified value $T_0$ ($t = t_0$). During ED-MD simulation, phonon and electron temperature profiles are established in the metal tip due to the Joule and Nottingham heating processes, while the temperature of bulk metal slab is kept as constant, i.e., $T = T_0$ (300 K). In the extensive bulk metal slab, heat flux in the lateral directions ($x$ and $y$ directions) normal to the box faces is set to 0. The Nottingham heat flux ($P_N$) is deposited to facet of finite element cell at the emitting surface of nanotip.

Finally, for the current flowing through the metal nanotip, the value is non-zero only in the tip structure (atomistic and coarse-grained sections). In the extensive bulk metal slab, both the electrical potential and current are set to zero. The total current passing in the metal tip satisfies the charge continuity condition, as given by Eq. (32).

$$\nabla \cdot (\sigma \nabla \varphi) = 0 \qquad (32)$$

The boundary condition associated with Eq. (32) on emitting surface is provided in Eq. (33), where the emission current density J is defined in Eq. (12).

$$\nabla \cdot (\sigma \nabla \varphi) = 0 \qquad (33)$$

## 4. Implementations and testing of FEcMD package

In this section, we would like to address the implementations and testing of algorithms in three main computational modules in FEcMD package, particularly focusing on the realization of the two-temperature heat conduction method in HC module, and the three-dimensional (3-D) field electron emission model corrected with space charge quantum effects. In addition, the issues related to the use of different numerical recipes and OpenMPI multithreads computational scheme on computational wall-times will be also briefly discussed.

### 4.1 Molecular dynamics simulation

FEcMD software package employs C++ to implement all algorithms related to classic molecular dynamics simulations. The fundamental principles and numerical algorithms of classic MD simulation can be found in Ref [[89]. Realization of MD simulation in FEcMD code is achieved through several key implementations and the corresponding subroutines including the thermodynamic ensembles (*ensemble.c*), neighbor lists (*nebrList.c*), atomic position and velocity integrations (*atomposition.c*), and statistics for structural and physical properties (*measurements.c*), as shown in Figure 2. MD and AF modules are required for running stand-alone molecular dynamics (MD) simulations in FEcMD program. The MD module calculates atomic accelerations and velocities to update the atomic positions according to Newton's second law. Meanwhile, the AF module provides various intrinsic and external forces acting on each atom in nanotips or nano-protrusions. Some key implementations and steps in the MD module are listed here.

- To achieve the high computational efficiency for a large system containing more than $10^5$ atoms, the inter-atomic forces are obtained using the neighbor-list algorithm [89]. This is achieved by updating the neighbor lists within a user defined cutoff radius ( $\Delta r$ ) and time interval ( $\sum(\max|v_i|) > \Delta r/2\Delta t$ ). The algorithm is included in the subroutine *nebrList.c*. Another two methods are also available in MD module to calculate forces among atoms, i.e., cell sub-division and all-pair methods [89]. All-pair approach is computationally expensive and highly inefficient for large atomistic models, compared to neighbor-list and cell sub-division algorithms.

- Both the leapfrog (See Eq. (34)) and prediction-correction (PC) methods are implemented for particle position and velocity integrations in the subroutine *atomposition.c*. The formulae used in PC method for atomic position and velocity can be found in Ref [89], and which are omitted here for brevity. Leapfrog integration method demands minimal storage space and less numerical calculation, which is an optimal option for large-scale MD simulation. Otherwise, prediction-correction algorithm shows higher numerical accuracy than that of leapfrog method, which is recommended for the rigid body and constrained MD dynamics.

$$\left. \begin{array}{l} \vec{v}_i(t+\Delta t/2) = \vec{v}_i(t-\Delta t/2) + \dfrac{1}{m_i}\vec{F}_i(t)\Delta t \\ \vec{r}_i(t+\Delta t) = \vec{r}_i(t) + \vec{v}_i(t+\Delta t/2)\Delta t \end{array} \right\} \qquad (34)$$

- Both the equilibrium and non-equilibrium molecular dynamics simulations are implemented in the MD module. For equilibrium MD simulation, microcanonical (NVE), canonical (NVT) and isothermal-isobaric (NPT) ensembles are supported in the subroutine *ensemble.c* [89]. In the case of non-equilibrium MD simulation, the MD module performs structural relaxation within a constant heat flux input in a specific direction, allowing users to calculate the lattice thermal conductivity of the materials using Fourier law. For equilibrium MD simulation, the three-dimensional periodic boundary conditions (PBCs) and Kubo-Greenwood formula are available to calculate the lattice thermal conductivity within equilibrium molecular dynamics simulation under NVE ensemble. The calculation of thermal conductivity is supported by the *heatdiffusion.cpp* subroutine in MD module.

- Supporting the machine learning interatomic potentials such as SNAP and MTP to conduct the

molecular dynamics simulations for multi-component alloys and compounds. Besides machine learning potentials, other widely used potentials including Lennard-Jones pair potential and EAM potential are also supported in MD simulations. The implementation of L-J potentials in FEcMD program directly follows the algorithms provided in Ref [[89]. Meanwhile, the EAM potential in FEcMD program is realized in using the open-source EAM potential library (*eam_alloy.c*) in LAMMPS code, mainly contributed by Foiles and coworkers [[89]. Regarding the machine learning potentials, SNAP in MD module is provided by the relevant library (*snap.c* subroutine) in LAMMPS program [[71],[72], while the realization of MTPs in FEcMD program is achieved by *mlip.cpp* library in the open-source machine-learning interatomic potential (MLIP) package [[75].

Here, we perform a series of molecular dynamics simulations using MD and AF modules in FEcMD software to validate the implemented methodology and algorithms. For such purposes, as shown in Figure 5, the conic Cu nano-tips with curvature radii of 1 nm, 5 nm, and 10 nm containing approximately $10^4$, $10^5$, and $10^6$ atoms are built, respectively. The standard molecular dynamics simulations are carried out for the three nanotips at 300 K using NVT ensemble and with a time step of 4 fs and a total duration of 1 ps. For all MD simulations, the lowest three atomic layers at the bottom of nanotips are fixed to their initial atomic positions. Otherwise, we place each nano-tip in a large rectangular box. The periodic boundary conditions are enforced in the lateral directions for the simulation box. Meanwhile, the box is aperiodic in the vertical z-direction. The EAM potential of bulk Cu metal is adopted to describe interatomic interactions among atoms in nanotips. The initialization of the velocities of all free atoms in nanotip follows the Maxwell-Boltzmann distribution. Updating the atomic accelerations, velocities and positions are realized using the PC/Leap-Frog time integration algorithms.

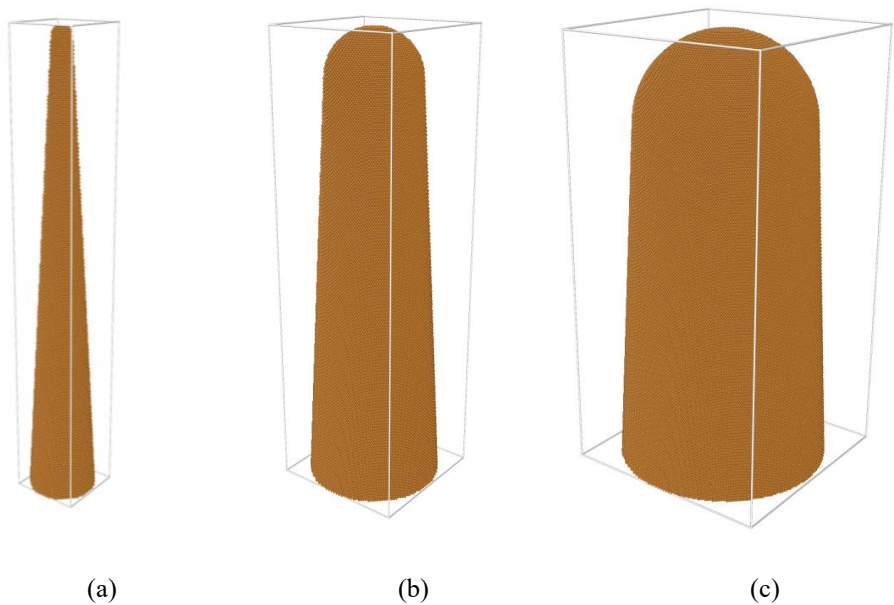

(a) (b) (c)

**Figure 5** Atomic structures of conic Cu nano-tips with a height $h_a$ = 50 nm, an opening angle $\theta$ = 3°, and curvature radii of (a) $r_0$ = 1 nm (76992 atoms); (b) $r_0$ = 5 nm (487976 atoms); (c) $r_0$ = 10 nm (1489049 atoms).

For molecular dynamics simulation employing the classic inter-atomic potentials of various forms (pair potentials, EAM, and machine learning potentials), the most time-consuming and memory demanding step is the calculation of the interatomic forces by counting atomic pairs at different distances. In the FEcMD software, three methods including neighbor-list, cell sub-division and all-pairs approaches are implemented in MD module. In Figure. 6, we demonstrate the scalability of the wall-time versus either the size of nanotips (total number of atoms) or the number of parallel threads using the three methods for 1 ps duration. As can be seen from Figure. 6(a), the neighbor-list method gives the best scalability for the total computational wall-time versus total number of atoms in nanotips, i.e., the computational time is always the shortest among all three methods regardless of the structure size. Nevertheless, all three methods perform similarly for small nanotips containing atoms less than $10^3$. The All-pairs method exhibits relatively poor scalability for the total wall-time versus nanotip size. For large nanotip (> $10^4$ atoms), the cell sub-division and neighbor-list methods follow the same scaling behaviors as shown in Figure. 6(a). Specifically, the all-pairs method counts all pairwise interactions in the system, resulting in a computational complexity of O($N^2$). This method is only suitable for small systems (<$10^3$ atoms), and as the number of atoms increases, the computation time increases significantly. For nanotips with ~$10^4$ atoms, the wall-time of the all-pairs method can be 50-150 times longer than the Neighbor-list or Cell division method, as shown in Figure 6(a). Both the cell sub-division and neighbor-list methods divide the system into small cells or groups to calculate the interatomic interactions inside the cells and within neighboring cells. This strategy greatly reduces the total number of atoms in the calculation for both methods, leading to the O($N$) scaling performance for computational wall-time versus nanotip size. Regarding the computational scalability versus the number of parallel threads of MD module in FEcMD software, the tests are carried out using the neighbor-list method, and the results are displayed in Figure. 6(b) for three different nanotips. The multi-thread calculation is realized using the open-source OpenMPI.1.7.3 library. All MD tests are conducted on a single Intel Xeon® CPU (E5-2680 v4 @ 2.40GHz) within 56 cores. For the smallest nanotip containing 76992 atoms, the use of multi-thread computation brings little gain in the wall times, compared to that of single-thread computation, as shown in Figure. 6(b). However, employing multi-thread computation is indeed critical to reduce the computational wall-time for large nanotips. From Figure. 6(b), it is obvious to see that the wall times do not scale linearly by increasing the number of CPU threads or cores in the task due to the increase of communication costs among processors. For the testing hardware employed here, the best wall-time saving is achieved using 8 CPU threads. Currently, FEcMD software does not support the multi-node parallel computing ability. The parallel scalability of MD module in FEcMD software on multi-node architectures is not performed.

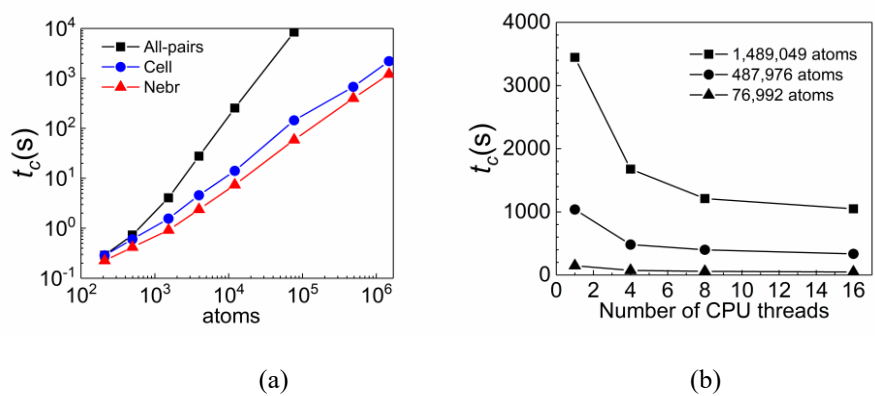

(a)          (b)

**Figure 6** Computational scalability of the algorithms (Neighbor-list, Cell sub-division, and All-pairs) employed in MD module to calculate the interatomic forces for a copper nano-tip ($h_a$ = 50 nm, $r_0$ = 1 nm, $\theta$ = 3°) using NVT ensemble at $T$ = 300 K for a total duration of 1 ps: (a) Wall-time consumption versus the total number of atoms in nano-tips; (b) Wall-time consumption versus number of parallel threads.

### 4.3 Two-temperature heat conduction simulation

The heat balance equations of phonons and electrons in two-temperature heat conduction simulation are given in Eqs. (27) and (28), respectively. In HC module of FEcMD package, we employ the rectangular finite different grids to numerically solve the head diffusion equations, and the discretization forms of phonon and electron heat balance equations are written as Eqs. (35) and (36).

$$C_\mathrm{p} \frac{T_\mathrm{p}^{i,j,k;t+\Delta t} - T_\mathrm{p}^{i,j,k;t}}{\Delta t} = G_\mathrm{ep}(T_\mathrm{e}^{i,j,k} - T_\mathrm{p}^{i,j,k}) +$$
$$\kappa_\mathrm{p} \left[ \frac{T_\mathrm{p}^{i+1,j,k} + p_\mathrm{p}^{i-1,j,k} - 2T_\mathrm{p}^{i,j,k}}{(\Delta x)^2} + \frac{T_\mathrm{p}^{i,j+1,k} + T_\mathrm{p}^{i,j-1,k} - 2T_\mathrm{p}^{i,j,k}}{(\Delta y)^2} + \frac{T_\mathrm{p}^{i,j,k+1} + T_\mathrm{p}^{i,j,k-1} - 2T_\mathrm{p}^{i,j,k}}{(\Delta z)^2} \right] \quad (35)$$

$$C_\mathrm{e} \frac{T_\mathrm{e}^{i,j,k;t+\Delta t} - T_\mathrm{e}^{i,j,k;t}}{\Delta t} = -G_\mathrm{ep}(T_\mathrm{e}^{i,j,k} - T_\mathrm{p}^{i,j,k}) + Q_k$$
$$\kappa_\mathrm{e} \left[ \frac{T_\mathrm{e}^{i+1,j,k} + T_\mathrm{e}^{i-1,j,k} - 2T_\mathrm{e}^{i,j,k}}{(\Delta x)^2} + \frac{T_\mathrm{e}^{i,j+1,k} + T_\mathrm{e}^{i,j-1,k} - 2T_\mathrm{e}^{i,j,k}}{(\Delta y)^2} + \frac{T_\mathrm{e}^{i,j,k+1} + T_\mathrm{e}^{i,j,k-1} - 2T_\mathrm{e}^{i,j,k}}{(\Delta z)^2} \right] \quad (36)$$

Where, the time step is given by $\Delta t$, and the grid spacings are denoted as $\Delta x$, $\Delta y$ and $\Delta z$ in $x$, $y$ and $z$ direction, respectively. The instantaneous phonon and electron temperatures at grid points are referred to $T_\mathrm{p}^{i,j,k;t}$ and $T_e^{i,j,k;t}$. In Eq. (36), $Q_k$ represents the heating sources such as resistive Joule heating and Nottingham heating processes, and which calculated by Eq. (37). Here, $A_k$ is the cross-section area of $k$th slab in the $z$-direction (or axial direction), while $J_i$, $S_i$ and $P_i$ are the emission current density, emitting surface area and Nottingham heating density of the same $k$th slab. The electrical resistivity for the sub-cell of $k$th slab in $z$-direction is given by $\rho_k$, and a linear interpolation scheme is used to estimate its value from the pre-tabulated electrical resistivity versus temperature data when the local sub-cell temperature is determined by the continuum heat conduction simulation.

$$Q_k = \frac{1}{A_k \Delta z} \left[ \left( I_{k-1} + \sum_{i \in k} J_i S_i \right) \rho_k \frac{\Delta z}{A_k} + \sum_{i \in k} P_i S_i \right] \quad (37)$$

Further simplification of the 3D heat balance equations is feasible, especially considering the metal protrusion has a large aspect ratio, and the local temperatures of sub-cells belonging to the same $k$th slab in the $z$-direction show no substantial differences. Then, the TTM may be solved only in the axial direction (or $z$-direction) by means of 1-D heat conduction process, assuming the local temperature is uniform in the radial direction for the $k$th slab. Under such an assumption, the corresponding discretization equations of heat balance equations for electrons and phonons are written by Eqs. (38) and (39), respectively.

$$C_e \frac{T_e^{k;t+\Delta t} - T_e^{k;t}}{\Delta t} = \kappa_e \left[ \frac{T_e^{k+1} + T_e^{k-1} - 2T_e^k}{(\Delta z)^2} \right] - G_{ep}(T_e^k - T_p^k) + Q_k \quad (38)$$

$$C_p \frac{T_p^{k;t+\Delta t} - T_p^{k;t}}{\Delta t} = G_{ep}(T_e^k - T_p^k) + \kappa_p \left[ \frac{T_p^{k+1} + T_p^{k-1} - 2T_p^k}{(\Delta z)^2} \right] \quad (39)$$

In our current implementations of Eqs. (35) - (39) for TTM, the specific heat of phonon subsystem ($C_p$) is approximated as a constant by Dulong-Petti law ($C_p$ = 3R), while the value of electron system is given by Eq. (30). Electron-phonon energy exchange rate ($G_{ep}$) is treated as a constant and which is specified by the simulated material.

For both 3-D and 1-D continuum simulations of electron and phonon heat conductions, the discretization equation in the time domain requires further evaluation. In HC module, the temperature profile of either electron or phonon in the continuum time domain is solved by the forward difference method, which is in general given as

$$\frac{\partial \vec{T}^n}{\partial t} \approx \frac{\vec{T}^{n+1} - \vec{T}^n}{\Delta t} \quad (40)$$

and the variation of local electron or phonon temperature from time $t$ to $t + \Delta t$ is approximated by

$$\vec{T}^{n+\Theta} = \Theta \vec{T}^{n+1} + (1-\Theta)\vec{T}^n \quad (41)$$

with the weighting factor $\Theta \in [0,1]$ in Eq. (41). The current implementation of Eqs. (35)-(39) for TTM adopts the fully implicit Euler scheme ($\Theta = 1$) in discretization equation of electron and phonon heat balance equations in continuum time domain. The fully implicit scheme is numerically more stable and robust than other two methods such as the fully explicit scheme ($\Theta = 0$) and Crank-Nicolson scheme ($\Theta = 0.5$), because the minimum time step in ED-MD simulation is restricted by the time step employed in the molecular dynamics simulation.

In both full-3D and simplified 1-D continuum heat conduction models, the rectangular 3D grid points divide the volume of the whole nanotip into smaller sub-cells, including both coarse-grained region and the upper atomistic model, as shown in Figure. 7. For each $k$th slab in the radial direction, surface FEM mesh points are assigned for calculating the total current density flowing through the nanotip and the total heat deposited to the electron subsystem using Eq. (37). In addition, each sub-cell in the atomistic domain is expected to contain at least dozens of atoms for performing a statistically meaningful velocity rescaling method that couples the continuum heat conduction temperature profile with the discrete particle dynamics in ED-MD methodology by Eq. (7).

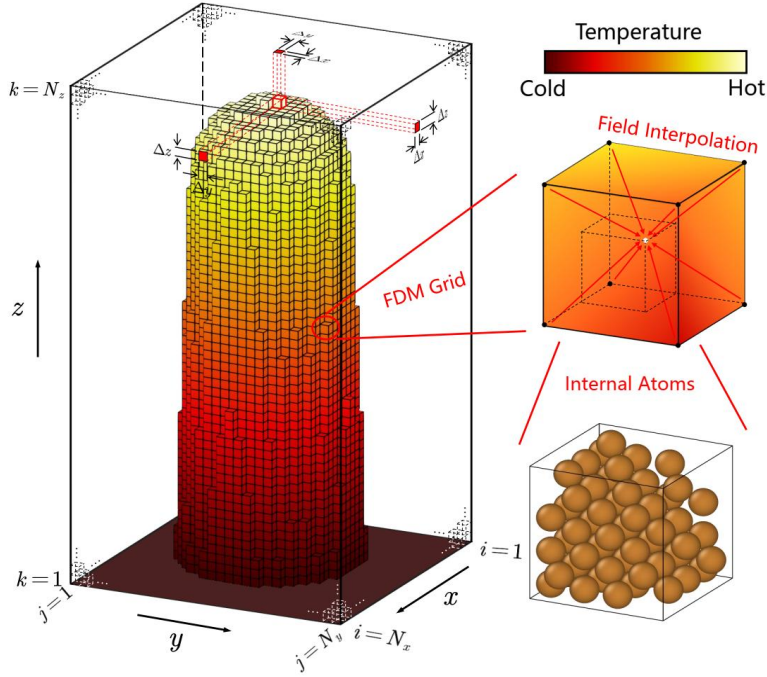

**Figure. 7** Multiscale rectangular finite difference grids in the continuum two-temperature heat conduction simulation coupled with molecular dynamics simulation in atomistic model using a local particle velocity rescaling scheme in each sub-cell. The rectangular mesh grid is given by $N_x \times N_y \times N_z$, with spacings $\Delta x$, $\Delta y$ and $\Delta z$ in x, y and z directions, respectively.

The implementations of TTM in HC module involve multiple steps as follows. The atomic structure of the nanotip is employed to define the atomistic domain through the coordination number analysis (CNA) for identifying all surface atoms at the boundary between the nanotip and vacuum, and as well as those between atomistic model and coarse-grained domain. This task is achieved using the locally modified *ProjectHeat.cpp* subroutine in FEMOCS library. Then, the rectangular volume ($l_x = x_{max} - x_{min}$, $l_y = y_{max} - y_{min}$ and $l_z = z_{max} - z_{min}$) defined by those surface atomistic atoms or coarse-grained atoms are used to generate finite difference grid points for metal nanotip. Note that the whole finite difference grid in continuum heat conduction simulation is automatically updated when the atomistic model has been changed in the MD simulation. This is mainly due to the use of a constant grid mesh in the continuum temperature simulation in HC module, i.e., $N_x \times N_y \times N_z$, as defined by the users in the input file. Therefore, the grid spacings such as $\Delta x$, $\Delta y$ and $\Delta z$ in Eqs. (35) and (36) are not fixed in heat conduction simulation, and which are determined from the instantaneous dimensions of nanotip volume ($l_x$, $l_y$ and $l_z$) and mesh parameters ($N_x$, $N_y$ and $N_z$). In fact, the atomistic model of nanotip is expected to be significantly elongated and deformed under the high E-field and strong heating processes shortly before the thermal runaway event, the larger grid spacings are needed to include more atoms in the sub-cells, resulting in the better statistical results for the velocity rescaling algorithm (See Figure. 7). Realization of TTM in continuum heat conduction simulation within the fully implicit scheme is freshly written and added to *ProjectHeat.cpp* subroutine in FEMOCS library. The calculation of Joule and Nottingham heats using Eq. (37) is implemented in *EmissionReader.cpp* subroutine which is integrated in our in-house *EleEmi.lib*. Finally, the HC module also outputs some important quantities such as instantaneous surface emission current density, the total emission current and electron and phonon temperature profiles by *ProjectHeat.cpp* subroutine in HC module.

For the benchmark tests of TTM heat conduction in HC module, we build a conical Cu nanotip containing 76992 atoms ($h = 50$ nm, $\theta = 3°$ and $r = 1$ nm). To initiate the resistive heating process (Joule heat) due to electron field emission mechanism, an external electric field of 250 MV/m is applied in all calculations using ED-MD-PIC scheme. The iteration time steps for MD, ED and HC modules are set as $\Delta t_{MD} = 4$ fs, $\Delta t_{PIC} = 0.5$ fs, $\Delta t_{HC} = 40$ fs, respectively. Therefore, rescaling of atomic velocities is performed for atomistic simulation in every 40 fs. Different slicing mesh grids for nanotips are tested including 3×3×100, 1×1×20 and 1×1×100. The temperature and size dependencies of electrical resistivity of Cu nanotip are obtained from previous works [[42]. The initial continuum and atomistic temperatures of the nanotip are set to 300 K. The continuum phonon temperature evolution and the effects of velocity rescaling scheme on the atomic kinetic energies of the atomistic model are illustrated in Figure 8. It is found that continuum phonon temperature of the nano-tip rapidly increases with the increasing of time steps, forming a temperature gradient of approximately 1000 K in the $z$ direction of the Cu nano-tip within the first 10 ps duration (See Figure 8(a)). The apex of nanotip is subjected to stronger resistive heating than the lower part. After rescaling velocities for atomistic model, local atomic kinetic energy shows a noticeable increase in the upper part of nanotip (See Figure 8(b)). Notably, rescaled atomic velocities still follow the Maxwell-Boltzmann distribution.

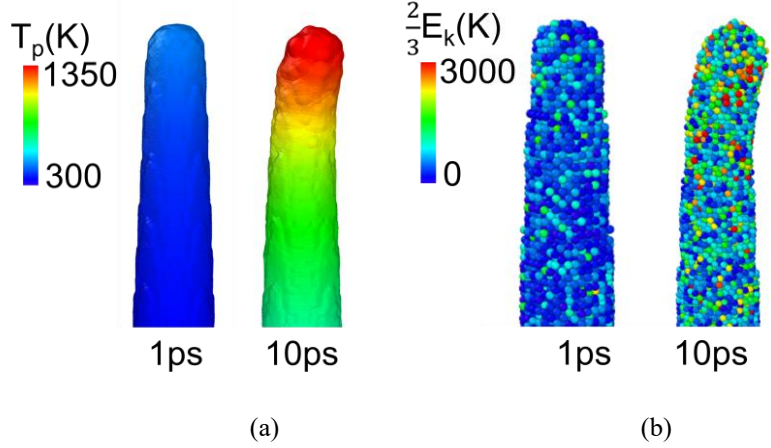

**Figure 8** Phonon temperature and atomic velocities rescaling in a hybrid ED-MD simulation for Cu nanotip containing 76992 atoms ($h_a$ = 50 nm, $\theta_0$ = 3° and $r$ = 1 nm): (a) Phonon temperature evolution; (b) Atomic kinetic energy before and after rescaling step. The cell dividing grid for temperature control is 3×3×100. The simulation is performed using 8 parallel threads on two CPUs (Intel Xeon® E5-2680 v4 @ 2.40GHz).

To quantify the accuracy and efficiency of temperature rescaling algorithm, we calculate the mean error (ME) in local temperatures between atomistic simulation and FEM results for sub-cells, as given by Eq. (42).

$$\text{ME} = \frac{1}{N_k} \sum_{j=1}^{N_k} \frac{T_k - \frac{2}{3}\frac{1}{N}\sum_{i=1, i \in k}^{N} E_{k,i}}{T_k} \tag{42}$$

Here, $N$ and $N_k$ represent the total number of atoms and total number of finite element mesh points in the same $k$th sub-cell, respectively; the local temperature of continuum mesh is given by $T_k$. First, we test the influence of the divided sub-cell size on the ME of temperature, and the results are shown in Figure 9(a) as a function of simulation time. When the continuum temperature distribution in the system tends to be uniform, such as when the instantaneous temperature is close to 300 K throughout the whole nanotip at the first several ps, using a dense mesh grid can result in small sub-cell volume. As a result, each sub-cell may contain few atoms (< 100). Thus, statistical error is signified in calculating the average atomistic temperature. As can be seen from Figure 9(a), the ME of 3×3×100 dividing grid is ~1% higher than those of 1×1×20 and 1×1×100 meshes in the first 2 ps. As the continuum temperature gradient in the Cu nanotip gradually develops with the increasing of MD duration, increasing the grid density in the axial direction is needed to reduce the ME that is caused by temperature non-uniformity and structural elongation. For example, the ME of 1×1×100 dividing grid is halved to that of 1×1×20 grid at 10 ps. Furthermore, the computational wall-time versus total number of dividing sub-cells is displayed in Figure 9(b). The total number of atoms contained in each cell is reduced by increasing the number of cells. Meanwhile, wall-time slightly increases with the number of sub-cells. For example, using 900 cells (3×3×100) costs 6.6% more computation wall-time than that of using 100 cells (1×1×100). In summary, dividing sub-cells in both axial and radial directions provides better numerical accuracy for large nanotip ($r > 10$ nm or micro-protrusion) than that of axial sub-cell only. Otherwise, the axial cell is sufficient for small nanotip ($r < 10$ nm). To be more specific, the characteristic dimensions of metal nanotip such as radius of curvature and the diameter in the radial direction are expected to be smaller than the phonon free mean path ($l_p$) for small tip, while the opposite conditions apply to the large tip. Nevertheless, users are allowed to specify the mesh parameters ($N_x$, $N_y$ and $N_z$) in the input file, and the influences on the atomistic model and other field electron emission properties can be tested and quantified.

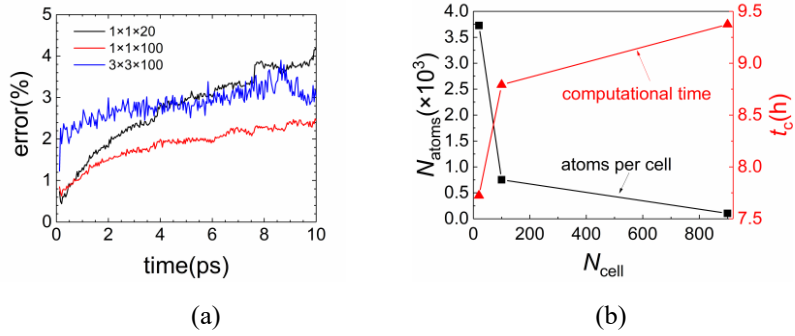

(a)  (b)

**Figure 9** Accuracy and computational efficiency of phonon temperature rescaling algorithm implemented in HC module: (a) Variations of the mean error (ME) with simulation time for different cell dividing grids; (b) Variations of the number of atoms per cell and total computation wall-time versus the number of cells. All calculations are conducted using 8 parallel threads on two CPUs (Intel Xeon® E5-2680 v4 @ 2.40GHz) processors for Cu nanotip containing 76992 atoms ($h_a = 50$ nm, $\theta = 3°$ and $r_0 = 1$ nm).

Finally, we employ the two-temperature thermal conduction model implemented in ED module to illustrate the decoupling of phonon and electron temperatures in the Cu micro-protrusion under the RF electric field. The RF electric field is defined by the angular frequency of 10 GHz and the amplitude of 300 MV/m. For Cu micro-protrusion is approximated by a conical stand terminated with a hemispherical cap. The total height and the radius of the hemispherical cap of Cu micro-protrusion are set to 2 μm ($h_0$) and 22.5 nm ($r_0$), respectively. Additionally, the conical tip has a half-aperture angle ($\theta_0$) of 3°. The thermophysical properties of electron and phonon of Cu are also needed to perform the two-temperature heat conduction simulation (See Eqs. (35)-(36)). All parameters employed in our simulations can be found in Ref [[54]. Notably, the lattice thermal conductivity of Cu is set to zero in the current study mainly because the phonon heat conduction is significantly smaller than that of electrons. Otherwise, the electric conductivity of Cu has also been corrected by the temperature and nano-size effects [[100-错误!未找到引用源。]. Since electron thermal conductivity of Cu nanotip is obtained from the Wiedemann-Franz law with the Lorentz number $L = 2.0 \times 10^{-8}$ W·Ω·K$^{-2}$, the electron thermal conductivity has the dependences on both the size and temperature of nano-tip [[102]. Otherwise, the electron-phonon coupling constant $G_{ep} = 4 \times 10^{17}$ W·m$^{-3}$·K$^{-1}$ [[103-[105]. The geometry of Cu micro-protrusion is rigid, and which is fixed to the initial shape during the ED simulation. The initial temperature of the Cu nanotip is set to 300 K and the total duration of the ED simulation lasts for 2 ns.

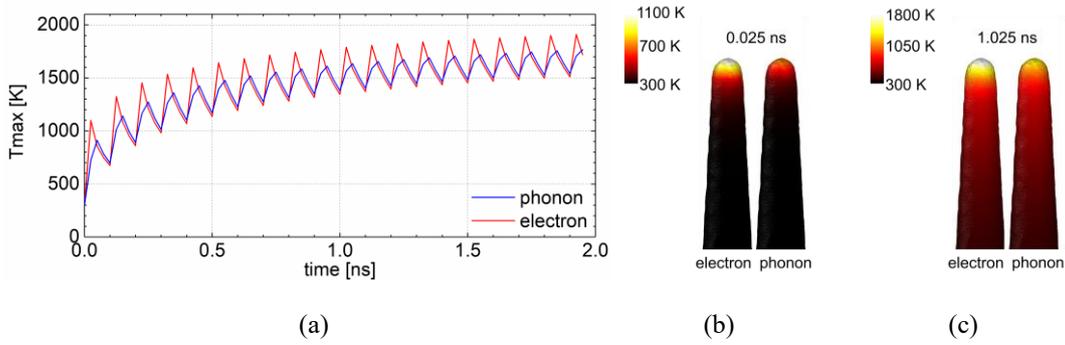

(a)     (b)     (c)

**Figure 10** The maximum temperature of electrons and phonons versus time at the apex of a conical Cu ($r_m$ = 22.5 nm, $h_m$ = 2 μm) micro-protrusion, respectively, under a radiofrequency electric field ($F$ = 300 MV/m, 10 GHz). (b) and (c) shows the corresponding electron and phonon temperature distributions of micro-protrusion at 0.025 ns and 2.025 ns, respectively.

The variations of phonon and electron temperatures versus time profiles are illustrated in Figure 10(a). The results clearly demonstrate the decoupling of electron and phonon temperatures in Cu micro-protrusion under RF electric field. More specifically, we observed the electron temperature is always higher than that of phonon during the first half-cycle (50 ps). Certainly, the electron emission process and the associated heating mechanisms (Joule and Nottingham heats) at the apex region of Cu micro-protrusion are synchronized with the periodicity of RF electric field (10 GHz). The duration of heating process now is comparable to that of electron-phonon relaxation time, and the decoupling of electron and phonon temperature evolutions are anticipated. It is also worth mentioning that the difference between phonon and electron temperatures inside the Cu micro-protrusion does not diminish in the entire simulation duration under the RF electric field. In Figures. 10(b) and 10(c), we also display the phonon and electron temperature distributions in the interior of Cu micro-protrusion at different simulation stages. Obviously, the electron and phonon temperatures show large difference at the apex region of Cu micro-protrusion in the beginning of simulation (0.025 ns). Meanwhile, the temperature distributions of electron and phonon subsystems are gradually synchronized at the late stage of ED simulation using the two-temperature model, indicating that the one-temperature model is indeed a valid approximation when either the electric field has the weak dependence on time or the on-set electric pre-breakdown time is sufficiently long, compared to that of electron-phonon relaxation.

**4.4 Field electron emission model with space charge quantum effects**

The overall workflow of the space charge quantum effects corrected field emission model implemented in ED module of FEcMD package is illustrated in Figure. 11. The emission current density is obtained for surface finite element cells of nano-tip after solving the 3-D Poisson equation to obtain local electric fields that are pointed toward the vacuum, adopting the multiscale hierarchical field electron emission scheme. In the proposed hierarchical emission model, electron emitting properties are evaluated using different levels of field emission theory for metal nanotip, including the classical Fowler-Nordheim equation (F-N), the WKBJ approximation and space charge quantum effects corrected field emission model (QMcFE). At the lowest level of emitting model, we differentiate the "blunt" surface domain (macroscopic domain) from the "sharp" domain (atomistic emission domain) using the sharpness parameter $\chi \equiv (\phi_{WF} - E)/eFR$ for the nanotip, where $E$, $F$ and $R$ represent the electron energy with respect to Fermi level, the local electric field value and the radius of curvature, as shown in Figure. 12. The critical value used to define the boundary between blunt and sharp emission regimes is the same as that of Ref [[60], i.e., $\chi_{max} = 0.1$. In the "blunt" regions ($\chi < 0.1$), the General Thermal-Field (GTF) equation is employed to calculate the field emission current density for the finite element cell on the surface of the nano-tip [[62]. Meanwhile, a full numerical calculation based on WKBJ model in sharp region is required to obtain the field emission current density. In the WKBJ model, the total electron tunneling potential $U(x)$ for each surface cell in the sharp region is calculated with and without space charge quantum effects, providing the "classical emitting regime" and the "quantum effects corrected emitting regime" as the second level of hagiarchy in the emitting model. Unlike the previous implementation of "classical emitting regime" where the space charge Coulomb potential ($\varphi_{sc}$) was not calculated, it is included in the classical regime in our emitting model by solving the 3-D Poisson equation in ED-PIC simulation in the continuum domain. The key criterion employed in ED module to define the boundary between classical and quantum emitting regimes is the ratio of emitting current densities with and without space charge quantum effects ($J_Q/J$), i.e., the quantum emitting regime is defined as $J_Q/J > 1.0$. In the classic regime ($J_Q/J <= 1.0$), emission current density of surface meshes is evaluated by WKBJ model, using the open-source GETELEC library. Initially, for estimating the ratio $J_Q/J$, the primary emission current density $J_Q$ is obtained by performing the non-self-consistent calculation of space charge density distribution and space charge exchange-correlation potential using the 1-D Poisson-Schrödinger equation in the electron emitting direction for all surface mesh facets in the sharp region, employing the boundary conditions that are defined by emission current density $J$ of classical emitting model. As can be seen from Figure. 11, the 1D coupled Poisson-Schrödinger equation is evaluated self-consistently to obtain the converged emission current density ($J = J_Q$), space charge density distribution ($n(x)$), exchange-correlation potential ($\varphi_{xc}(x)$), space charge Coulomb potential ($\varphi_{sc}(x)$), image charge potential ($\varphi_{ic}(x)$) for quantum emitting regime [[56[58]. Here, we would like to emphasize that the proposed 1D electron emitting model is applied to all surface cells satisfying $J_Q/J > 1$, and only in those surface areas, space charge quantum effects are expected to significantly alter both the shape and height of electron tunneling energy profile and thus changing the emission current density. Meanwhile, for other remaining surface meshes, the local electric field ($\varphi_{sc}(x)$) and space charge distribution are determined from the usual three-dimensional ED-PIC simulation [[45].

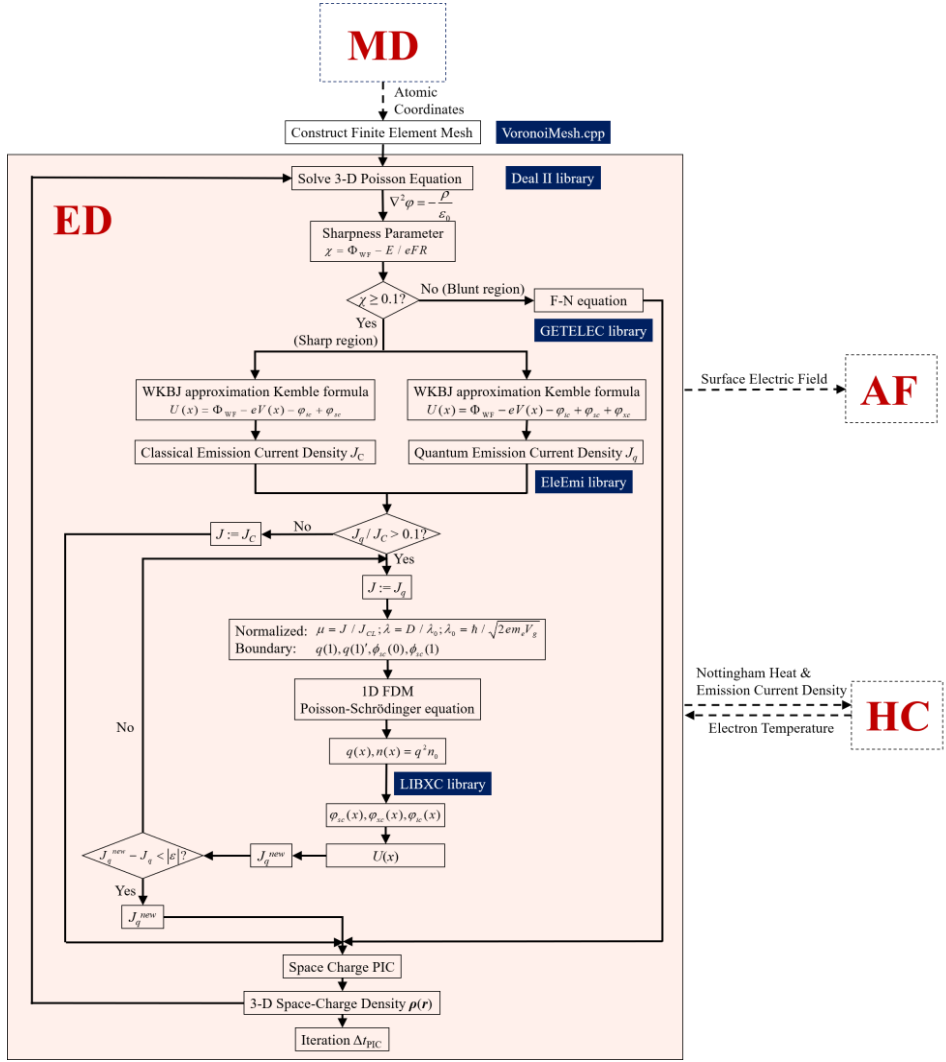

**Figure 11** Workflow and key algorithms of electrodynamics module (ED) for calculating electric field evolution and electron field emission characteristics. The principal ED module is highlighted as the solid square. Other supporting modules are denoted by dashed squares including MD, AF and HC modules.

For quantum regime ($J_Q/J > 1.0$), the discretization equations of Eq. (22) within the finite difference method are given by

$$\frac{q_{i+1} + q_{i-1} - 2q_i}{\Delta x^2} + \frac{\lambda^2}{D^2}\left[\frac{\varphi_{\text{sc}i} - \varphi_{\text{xc}i}}{eV_g} + \frac{x}{D} - \frac{4}{9}\frac{\mu}{q_i^4}\right]q_i = 0 \quad (43)$$

$$\frac{\varphi_{\text{sc}i+1} + \varphi_{\text{sc}i-1} - 2\varphi_{\text{sc}i}}{\Delta x^2} = \frac{2}{3}\frac{eV_g}{D^2}q_i^2 \quad (44)$$

here, $i$ denotes the discrete grid points, and other physical quantities are already defined in Eqs. (22)-(26). Boundary conditions adopted here for self-consistently solving Eqs. (43) and (44) to obtain space charge wave amplitude ($q$) and electric field in the electron emitting direction are given as $q(1) = (2\mu/3)^{1/2}$, $q'(1) = 0$, $\varphi_{sc}(0) = 0$ and $\varphi_{sc}(1) = 0$ (Dirichlet boundary conditions for static applied E-field) or $\varphi'_{sc}(0) = s_c$ and $\varphi'_{sc}(1) = s_a$ (Neumann boundary conditions for radiofrequency E-field). In Eq. (43), the normalized current density $\mu$ is updated using the fully implicit scheme as $\mu^{n+\Theta} = \Theta\mu^{n+1} + (1-\Theta)\mu^n$ with $\Theta = 1$.

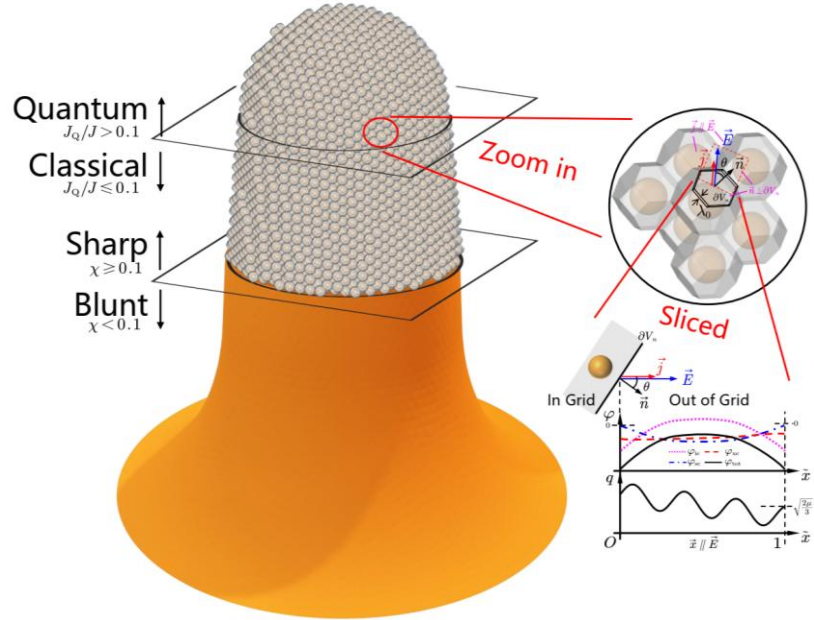

**Figure 12**. The schematical illustration shows the implementation of a multiscale hierarchical field electron emission model in the FEcMD program involving the space charge quantum effects. The continuum and atomistic domains of metal nanotip are indicated as the smooth surface and granular apex, while solving the 1-D Poisson-Schrödinger equation for each surface mesh facet is also highlighted.

The subroutines and libraries relevant to the implementation of space charge quantum effects corrected field electron emission model are as follows: the creation of atomistic model, coarse-grained domain (lower section of tip and the bottom bulk slab) for metal nanotip is realized in *tipMaker.c* subroutine, supporting several different geometries (See Appendix A) for tip apex; the 3-D multi-scale self-adaptive finite element mesh for the continuum domain is generated using the *VoronoiMesh.cpp* subroutine in FEMOCS library; solving the 3D Poisson equation for electric potential using FEM is achieved by Deal.II library; the space charge distribution in the vacuum is simulated using *Pic.cpp* subroutine in FEMOCS library; the GETELEC library is employed to calculate emission current density from classical Fowler-Nordheim equation or WKBJ model for blunt region or $J_Q/J < 1$ emitting regime;

solving the 1-D Poisson-Schrödinger equation for the space charge quantum effects corrected field electron emission model is implemented in the in-house *EmissionReader.cpp* subroutine; computing the exchange-correlation potential of space charge is realized using the open-source Libxc library. Note that all aforementioned subroutines and open-source libraries are fully integrated into ED module of FEcMD package.

In Figure. 12, several critical parameters are highlighted for the current implementation of 1-D coupled Poisson-Schrödinger equation in the proposed quantum emitting model, i.e., the gap spacing $D$, the number of grid points ($N_{1D}$), and the normalized field emission cutoff distance ($\alpha = LF/\Phi_{WF}$). As noted previously, the gap spacing $D$ appearing in the original 1-D coupled Poisson-Schrodinger equation defines the distance between two planar electrodes in the nanogap [[58,[59]. In fact, for the planar electrode configuration with nanogap and under the Dirichlet boundary conditions, the applied macroscopic E-field is defined as $E = V_g/D$. However, for 3-D nano-tip, the parameter $D$ serves as an auxiliary input parameter in the electron emitting direction for each surface mesh facet when solving the Poisson-Schrödinger equation, and which must be specified by the users in the input file (the tag: *gap_distance* in Appendix C) under Dirichlet boundary conditions for electric field. Meanwhile, there is no need to specify the value of parameter $D$ by the user within the Neumann boundary conditions, i.e., $\varphi'_{sc}(0) = s_c$ and $\varphi'_{sc}(1) = s_a$. Regarding the total number of grid points used for the self-consistent 1-D Poisson-Schrödinger equation, the recommended value has been determined to be 200 in our previous work, representing a good tradeoff between computational efficiency and numerical accuracy [[59]. Nevertheless, the number of grid points ($N_{1D}$) can be given by the users using the tag *n-lines* in the input file of ED module (See Appendix C).

The last remaining parameter that needs to be tested is the local field emission cutoff distance $\alpha$ (Tag: *alpha* in Appendix C). The cutoff distance indicates the furthest boundary alone the emitting direction in solving the 1-D Poisson-Schrödinger equation, and which critically affects the spatial window for evaluating the electron tunneling energy barrier profile. For such a purpose, we carry out ED-MD simulations using the Cu nano-tip ($h_0 = 50$ nm, $r_0 = 1$ nm, $\theta_0 = 3°$), as shown in Figure 5(a) as the cathode, with the tip apex positioned 10 nm away from the anode, and a voltage of 20 V applied. The exchange-correlation effects are treated at the level of generalized gradient approximation (GGA) using the Perdew-Burke-Ernzerhof (PBE) exchange-correlation functional [104,[105]. The ED-MD simulation is conducted for 4 fs to fully relax the atomistic model of Cu nanotip with a time step of 1 fs. Then, we solve the 1-D Poisson-Schrödinger equation using different local cutoff distances under WKBJ model. The impact of different normalized local cutoff distances $\alpha = LF/\Phi_{WF}$ on the maximum emission current density $J_{max}$, electron density profile $n(x)$, exchange-correlation potential profile $\varphi_{xc}(x)$, and tunneling barrier profile $U(x)$ in the electron emitting direction at the tip apex are depicted in Figure 13. As shown in Figure 13(a), when $\alpha$ is less than 1.5, the maximum emission current density and average exchange-correlation potential profiles do not show oscillating behaviors. This is because the local cutoff distance is too small, and which approaches the electron de Broglie wavelength, as shown in Figures 13 (b) and 13(c) ($\alpha = 1$). In addition, the tunneling barrier profile is not fully reproduced (see Figure 13 (d)) within a too small local cutoff distance, introducing large numerical errors in the calculation of the field emission tunneling probability. When we set $\alpha > 1.5$, both the emission current density and exchange-correlation potential profiles oscillate with the increasing of $\alpha$ values with the well-defined mean values, indicating the numerical convergence is reached for the given $\alpha$. Besides Cu with face-centered cubic lattice structure, nanotips consisting of Mo, W and Ti are also tested by varying the local emission cutoff distance. From those tests, it is concluded that when the normalized local cutoff distance $\alpha$ is set to a value larger than 2.0, the self-consistent solutions of the Poisson-Schrödinger equation in the quantum electron field emission regime is numerically accurate and robust. It is also worth mentioning that the electron tunnelling energy profiles shown in Figure 13(d) have finite values on the emitting surface, because the image charge potential is calculated using Eq. (21) under the theoretical framework of Thomas-Fermi free electron model and RPA electron screening approximation [[101].

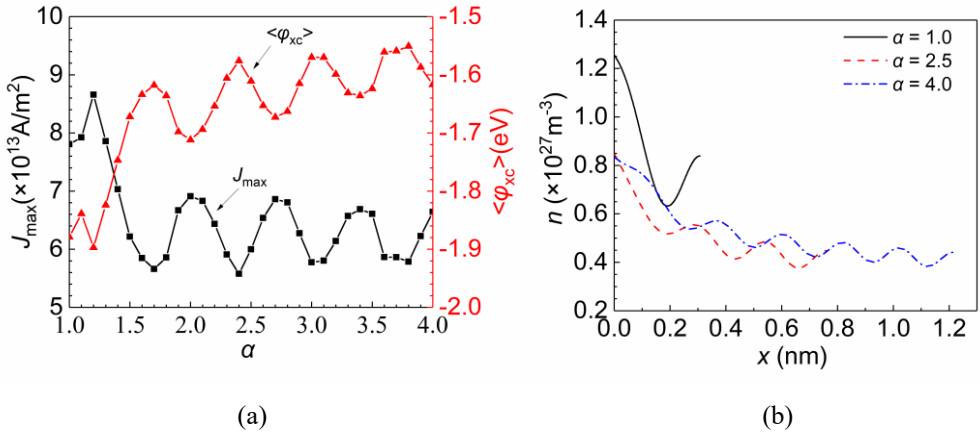

(a)　　　　　　　　　　　　　(b)

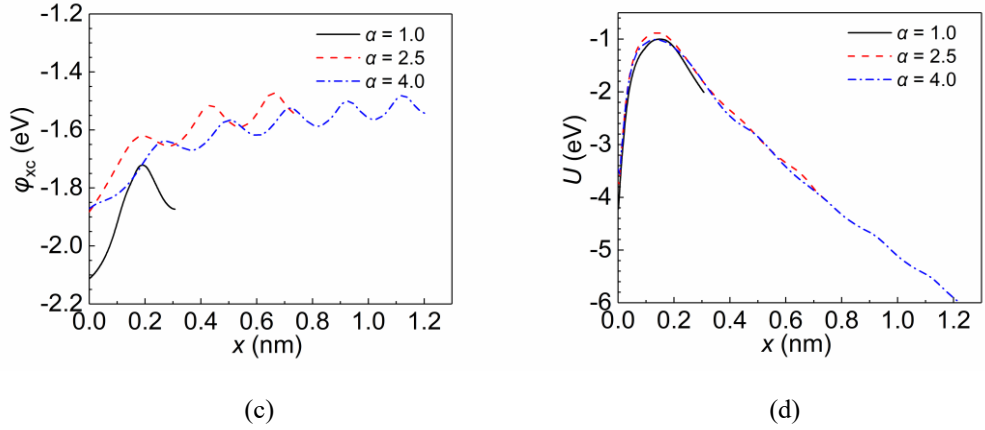

|     |     |
| :-: | :-: |
| (c) | (d) |

**Figure 13** Tests of solving the Schrödinger equation and the quantum-modified field emission model with various localized distances on the Cu nano-tips ($h_0$ = 100 nm, $r_0$ = 1 nm, $\theta_0$ = 3°) as the cathode, with the tip apex positioned 10 nm away from the anode, and a voltage of 20 V applied ($F_{max}$ = 15.1 GV/m) for ED-MD simulations until the current stabilizes (4 fs): (a) The maximum current density $J_{max}$ and average exchange-correlation potential $<\varphi_{xc}>$ at the tip apex as a function of the normalized localized distance $\alpha = LF/\Phi_{WF}$ used in solving the Schrödinger equation; (b) The localized electron density profile, (c) exchange-correlation potential profile $\varphi_{xc}(x)$, and (d) tunneling barrier profile $U(x)$ in the vertical direction at the tip apex are displayed for $\alpha$ = 1.0, 2.5 and 4.0, respectively.

In Figures 14(a) and 14(b), the 3-D distributions of local mean space charge density and average exchange-correlation potential on the emitting surface are shown under the applied E-field of 320 MV/m for Cu nanotip ($h_0$ = 100 nm, $r_0$ = 1 nm, $\theta_0$ = 3°). Obviously, there is a strong correlation between the electron density and exchange-correlation potential profiles. To be more specific, electron emission is enhanced by the presence of exchange-correlation potential in the proposed quantum emitting regime, mainly because the space charge quantum effects could lower the electron tunneling barrier height [[58],[59]]. Due to the utilization of the Poisson-Schrödinger equation to describe the wave amplitude of space charge density distribution under a linear external potential, the electron density near the nano-tip surface shows the distinct wave-like oscillation (Airy functions).

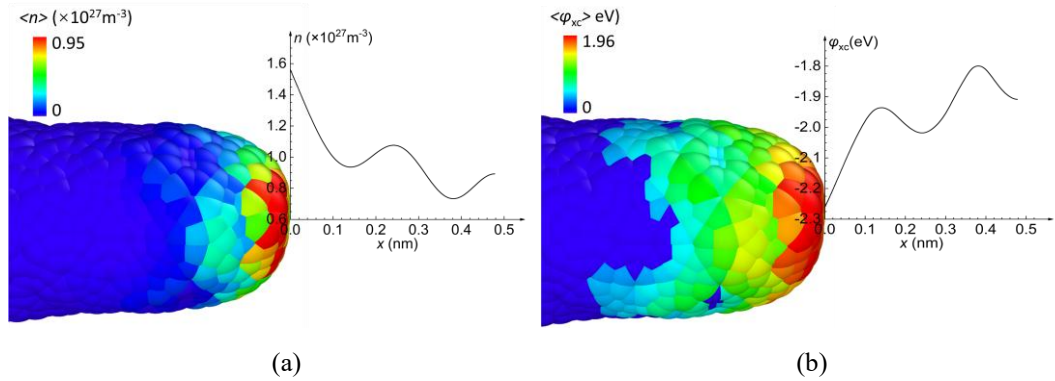

(a)                                     (b)

**Figure 14** The 3-D distribution of (a) the average electron density and (b) the average exchange-correlation potential near the surface for a Cu nano-tip ($h_t$ = 100 nm, $r_0$ = 1 nm and $\theta_0$ = 3°; $F$ = 320 MV/m) at 4 fs. The corresponding electron density profile and the exchange-correlation potential profile within the localized distance at the tip apex are also given in the insets.

**4.5 Finite element simulations for electric fields and temperature**

HC and ED modules in FEcMD program provide stand-alone FEM-based solvers to perform the advanced multi-physics simulations for micro- and nano-protrusions or electron emitters. This feature is very useful for a quick evaluation of the electron emission characteristics of metal electron emitter when the physical properties and geometry are specified. FEcMD program relies on a built-in subroutine (*tipMaker.c*) to generate various geometries and sizes for micro- and nano-protrusions, including hemisphere, cylinder, pyramidal-hemisphere [[26]], prolate-spheroidal [错误!未找到引用源。], hemi-ellipsoidal [错误!未找到引用源。] and mushroom-head [[27]]. Each geometry is specified by a set of parameters in the relevant scripts. FEcMD program provides those scripts for users to build tip structures for their own research interests. . Then the atomic structure with the user defined geometry is imported into the FEMOCS library to produce the finite element mesh for multi-physics simulation. In Figure A1 and Table. A1 of Appendix A, the supported geometries and their featured parameters are summarized. Additionally, an example of the script for generating various tip geometries is also provided in Appendix A.

Here, we conduct multi-physics simulations for three different electron emitter geometries using FEcMD program, and the results are illustrated in Figure 15. In Figures 15(a)-(c), the electric field vectors and temperature distribution are shown for prolate-spheroidal, hemi-ellipsoidal and mushroom-head emission tips at different simulation times, respectively. Those three geometries exhibit a relatively sharp apex region, resulting in high local electric field strength and strong electron emission process. Without invoking the MD module, the FEcMD software can be employed to study the field emission properties and even the electrical pre-breakdown E-field for micro-emitter or micro-protrusions. By performing the standard finite-element simulation within ED and HC modules in FEcMD software, the pre-breakdown E-field or the time to thermal runaway can be estimated from the calculated phonon temperature of electron emitter.

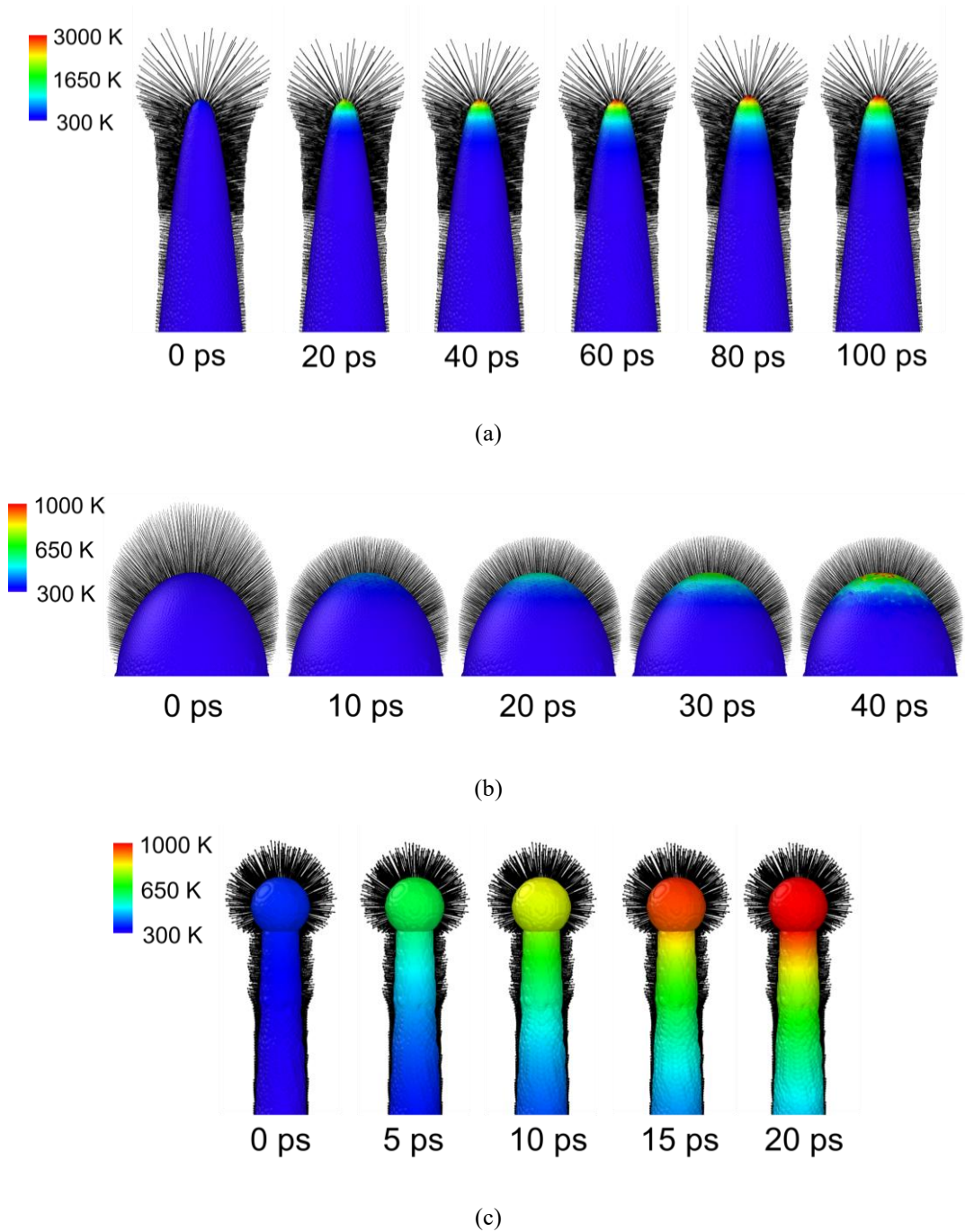

(a)

(b)

(c)

**Figure 15** The multi-physics simulations of electron emitters with various geometries for electric field and temperature distributions: (a): the prolate-spheroidal emission tip ($r$ = 5 nm, $h$ = 1 μm, $\theta$ = 4°) under a constant electric field (600 MV/m); (b): the hemi-ellipsoidal emission tip ($r_b$ = 1.5 μm, $h$ = 2.5 μm) under a constant electric

field (4 GV/m). (c): the mushroom-head emission tip ($r_t$ = 1.5 nm, $R$ = 3 nm, $h$ = 100 nm, $\theta$ = 3°) under a constant electric field (400 MV/m). The parameters of a specific tip geometry are defined in Appendix A.

# 6. Conclusions

In this paper, we presented the methodologies, implementations and critical testing of FEcMD software package, a multi-physics and multi-scale simulation tool for investigating and understanding atomic structure evolutions, phase transitions and electron emission characteristics of micro- and nano-protrusions under electric filed and heat conduction. FEcMD program offered several key upgrades in the algorithms that are outperform the existing multiscale-multiphysics ED-MD simulations, including the two-temperature heat conduction simulation, quantum effects corrected field electron emission model (exchange-correlation effects and the image charge potential within TFA), and the interface to support machine learning potentials for multi-component alloys. We believe those major upgrades available in FEcMD software package could not only provide a powerful computational methodology to study deformation and phase transition of metal micro- or nano-protrusions with electric field and heat, but also greatly benefits the design of field electron emission based modern nanoelectronics and nano-emitters.

**Acknowledgements**

This research is financially supported by the Young Talent Support Plan at Xi'an Jiaotong University awarded Bing Xiao (No: DQ1J009) and the Fundamental Research Funds of the Central Universities (No. xtr052024009 and No. xyz022023092). Bing Xiao would like to thank Prof. Flyura Djurabekova (Helsinki Institute of Physics and Department of Physics, University of Helsinki), Prof. Andreas Kyritsakis (Institute of Technology, University of Tartu), Dr. Mihkel Veske (Helsinki Institute of Physics and Department of Physics, University of Helsinki) for fruitful discussions at the early phase of the project.

**Appendix A: Create different geometries of nanotips**

The FEcMD software provides scripts for the users to build the nanotips with six different geometries, including Pyramidal-hemisphere, Mushroom-head tip, Prolate-spheroidal, Hemi-ellipsoidal, Cylinder and Cone. In Figure A1 and Table A1, the schematic diagram of all geometries along with the corresponding analytical expressions are provided.

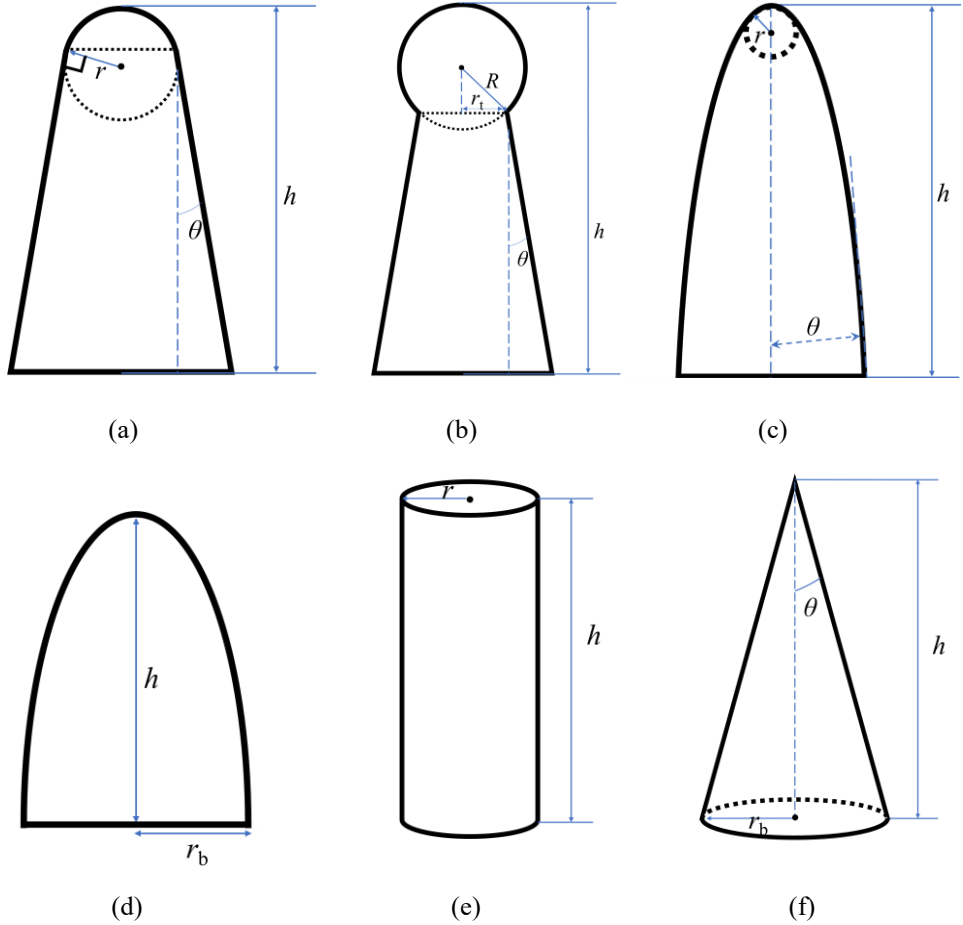

**Figure A1** Schematic diagrams of various types of emitter tips that can be built by the MD module and parameters determined by users: (a): the pyramidal-hemisphere emission tip; (b): the mushroom-head emission tip; (c): the prolate-spheroidal emission tip; (d): the hemi-ellipsoidal emission tip; (e): the cylinder emission tip; (f): the cone emission tip.

**Table A1** Types of emitter tips that can be built by the MD module using *tipMaker.c* subroutine, corresponding expressions and parameters determined by users.

| Type | Expression | Description |
|---|---|---|
| (a) Pyramidal-hemisphere | $\begin{cases} x^2+y^2+(z-h+r)^2 = r^2, & z \geq h-r(1-\sin\theta) \\ \sqrt{x^2+y^2}+z\tan\theta = \\ r\cos\theta+\tan\theta[h-r(1-\sin\theta)], & 0 \leq z < h-r(1-\sin\theta) \end{cases}$ | Total height $h$, Radius of the hemispherical cap $r$, Half-angle $\theta$ |
| (b) Mushroom-head tip | $\begin{cases} x^2+y^2+(z-h+R)^2 = R^2, & z \geq h-R-\sqrt{R^2-r_t^2} \\ \sqrt{x^2+y^2}+z\tan\theta = \\ r_t+\tan\theta\left(h-R-\sqrt{R^2-r_t^2}\right), & 0 \leq z < h-R-\sqrt{R^2-r_t^2} \end{cases}$ | Total height $h$, Radius of the mushroom cap $R$, Radius of cone top $r_t$, Half-angle $\theta$ |

| Type | Expression | Description |
|---|---|---|
| (c) Prolate-spheroidal | $\begin{cases} x = a\sinh(u)\sin(v)\cos(\varphi) \\ y = a\sinh(u)\sin(v)\sin(\varphi) \\ z = a\cosh(u)\cos(v) \end{cases}$, $\begin{pmatrix} 0 \le u < u_{max} \\ 0 \le \varphi < 2\pi \end{pmatrix}$<br><br>the half of the foci distance $a = \dfrac{r}{\sin\theta \tan\theta}$,<br><br>$v = \pi - \theta$,<br><br>$t = \dfrac{a\cos(v) - h}{a\cos(v)}$, $u_{max} = \log\left(t + \sqrt{t^2 - 1}\right)$,<br><br>the distance from the tip top to anode $d = a\cos\theta$ | Total height $h$,<br>Tip radius $r$,<br>Half-angle $\theta$ |
| (d) Hemi-ellipsoidal | $\begin{cases} x = r_b \sin(\theta)\sin(\varphi) \\ y = r_b \cos(\theta)\sin(\varphi) \\ z = h\cos(\varphi) \end{cases}$, $\begin{pmatrix} 0 \le \theta < 2\pi \\ 0 \le \varphi < \dfrac{\pi}{2} \end{pmatrix}$ | Total height $h$,<br>Radius of the bottom $r_b$ |
| (e) Cylinder | $x^2 + y^2 = r$, $0 \le z \le h$ | Total height $h$,<br>Radius of the bottom and the top $r$ |
| (f) Cone | $\sqrt{x^2 + y^2} + z\tan\theta = r_b$, $0 \le z \le h$<br><br>$r_b = h\tan(\theta)$ | Total height $h$,<br>Radius of the bottom $r_b$ |

To obtain the atomic coordinates file *mdlat.in.xyz* for each geometry of emission tip, user needs to provide two input files separately in a target folder:

(1) Unit cell of the bulk metal in the file *CONTCAR* for building a specified tip geometry.

(2) Tip parameters in the input file *md.in*.

The *CONTCAR* file defines the lattice structure of bulk metal in terms of lattice vectors, atomic coordinates, types of elements and number of atoms, and its format is precisely the same as that of the file employe in standard VASP code.

**Appendix B. Input and output files for molecular dynamics.**

To run the stand-along molecular dynamics simulation, the MD and AF modules need to be executed simultaneously. Before running a simulation, several input files should be created with the proper designated keywords to initiate the calculation, and most relevant and important instantaneous and final results are written in multiple output files during the simulation. Figure B1 shows a typical directory structure for input and output files in a single MD simulation task. Usages of both input files and output files appear in Figure B1 are explained in Table B1. Furthermore, detailed settings of key-value pairs in file *md.in* for starting a MD calculation are listed in Table B2.

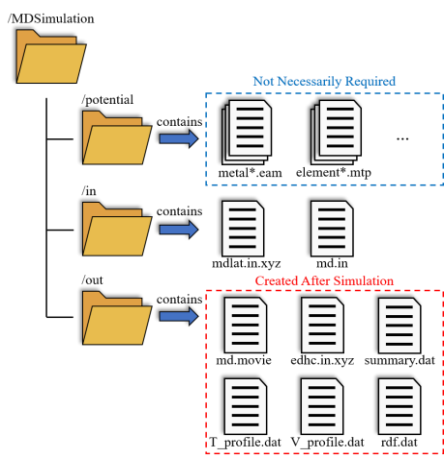

**Figure B1** Typical directory structure for both input and output files in a stand-alone MD simulation.

**Table B1** List of input files and output files for MD simulation. Contents of these files are also declared.

| File | I/O | Description |
| --- | --- | --- |
| metal*.eam | I | The file for interatomic potential, supporting types are declared by the tag: *force_type*. |
| mdlat.in.xyz | I | The initial atomic coordinates for the tip. |
| md.in | I | The input parameters for MD simulation. |
| md.movie | O | Trajectory of atomistic model during MD. |
| edhc.in.xyz | O | Instantaneous atomic coordinates exported to ED and HD modules. |
| summary.dat | O | The average energy, average temperature, tip height for atomistic model. |
| T_profile.dat | O | The temperature profile obtained from MD simulation. |
| V_profile.dat | O | The instantaneous atomic velocities for atomistic model. |
| rdf.dat | O | Results for radial distribution function (RDF). |

**Table B2** List of keywords in file *md.in* for starting MD simulation. The variable, type, default value, unit and description are shown in detail for each keyword. (Notations for type: R for real type, I for integer type and S for string)

| Keyword | Type | Default | Unit | Description |
| --- | --- | --- | --- | --- |
| deltaT | R | 4.0 | fs | Time step. |
| temperature | R | 300.0 | K | Initial temperature. |
| Nelems | I | 1 | — | The number of elements. |
| elems | S | Cu | — | Element types. For multiple elements, all elements should be listed with a spacing among them, i.e., W Mo |
| mass | R | 63.546 | u | Relative atom mass for each element. For multiple elements, all values must be provided in the same order |

| Keyword | Type | Default | Unit | Description |
|---|---|---|---|---|
| | | | | as that of element symbols with a spacing among them, i.e., 183.84 95.94. |
| stepAvg | I | 10 | — | The time interval for saving energy and temperature. |
| stepMovie | I | 100 | — | The time interval to output atomic trajectory. |
| stepEquil | I | 1000 | — | Total time steps to equilibrate the atomistic model. |
| stepLimit | I | 100000 | — | Total number of simulation steps. |
| randSeed | I | 7 | — | Random seed. |
| boundary | S | p | — | Boundary condition: periodic (p) or nonperiodic (n). |
| structure_type | S | FCC | — | Lattice structure type: FCC, BCC and cubic. |
| lattice | R | 3.6147 | Å | Lattice parameters, $a$. |
| interact_method | S | nebr | — | Methods to compute interatomic interactions: allpairs, cell or nebr. |
| force_type | S | MTP | — | Types of atomic interactions: lj, metal, alloy, snap, eamfs, mtp. |
| force_file | S | ./Cu.mtp | — | Atomic potential file name. |
| ensemble | S | NVT | — | Ensembles: NVE, NVT, NPT or nonEquil. |
| Nthreads | I | 8 | — | Number of threads for MD in parallel computing task. |
| pedestal_thick | I | 0 | — | Fixed atomic coordinates at the bottom for atomistic model. |
| cell_order | S | disorder | — | Types of atomic model created for alloys: order or disorder. |
| initUcell | R | 10 10 10 | — | The supercell dimensions with respect to unit cell for creating the tip structure; the dimensions must be larger than the largest diameter of tip in radial directions, and the height of tip ($h_a$) in the atomistic domain in the axial direction. |
| tip_file | S | ./mdlat.in.xyz | — | Input file for atomic coordinates of nanotip. |
| tip_R | R | 30.0 | Å | The radius of the mushroom cap $R$. |
| tip_r | R | 15.0 | Å | The radius of curvature of conical tip $r_0$. |
| tip_h | R | 500.0 | Å | Total height of atomistic model of tip $h_a$. |
| tip_theta | R | 3.0 | ° | Half-angle of tip $\theta_0$. |

**Appendix C. Input and output files for ED+HC simulations.**

Running the stand-alone electrodynamics (ED) and heat conduction (HC) modules allow for calculating electric fields, field emission, and heat conduction in three-dimensional rigid nano-tips using the finite element method. To execute these simulations, users must prepare specific input files defining the mesh geometry, computational parameters, and material properties. Figure C1 illustrates the typical directory structure for an ED+HC simulation. The functions of both input and output files are detailed in Table C1.

Additionally, Table C2 provides a comprehensive list of keywords used in the input file *edhc.in* to set the simulation parameters.

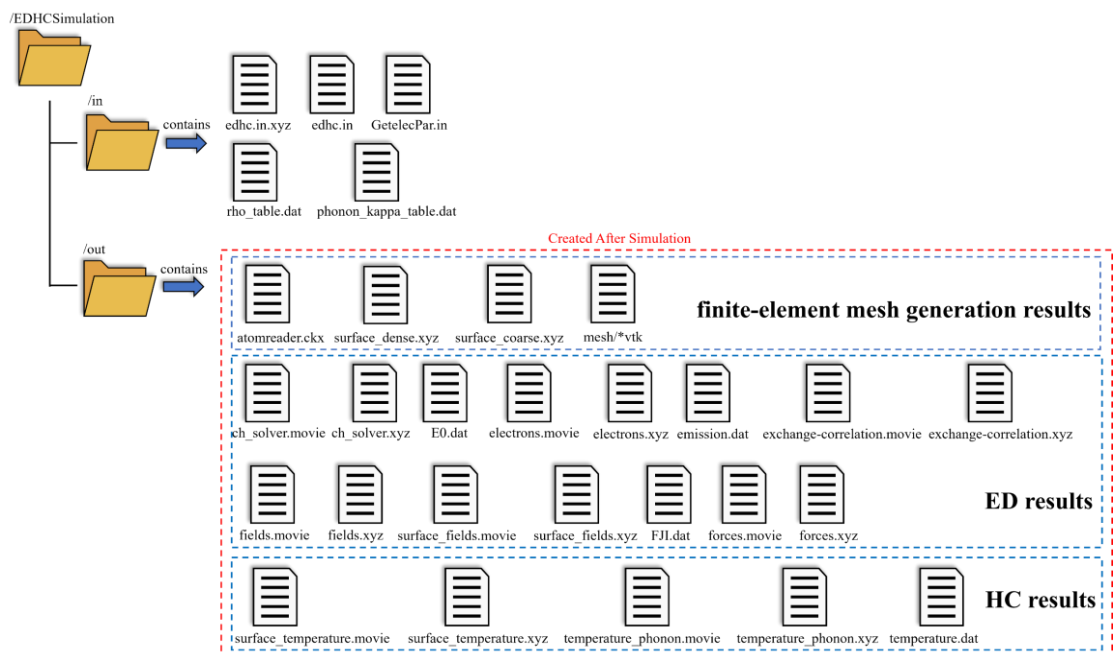

**Figure C1.** Typical directory structure for both input and output files in a standard ED+HC simulation.

**Table C1.** List of input files and output files for ED+HC simulations. Contents of these files are also declared.

| File | I/O | Description |
|---|---|---|
| edhc.in.xyz | I | The instantaneous finite element mesh file. |
| edhc.in | I | Input file for simulation parameters. |
| GetelecPar.in | I | The auxiliary input file required by GETELEC.lib. |
| rho_table.dat | I | The electrical resistivity with "rho_table" format. |
| phonon_kappa_table.dat | I | The phonon thermal conductivity file. |
| atomreader.ckx | O | The atomic coordinates read by FECMOS library. |
| surface_dense.xyz | O | The mesh coordinates of tip surface points in the atomistic domain. |
| surface_coarse.xyz | O | The mesh coordinates of tip surface points in the coarse-grained domain. |
| mesh/*vtk | O | Meshes generated for finite element calculation. |
| ch_solver.movie | O | Saved electric current and heat conduction simulation results of all previous steps. |
| ch_solver.xyz | O | Saved electric current and heat conduction simulation results of the current step. |
| E0.dat | O | Values of the applied time-dependent electric field for previous steps. |
| electrons.movie | O | Space charge (Emitting electrons) trajectory obtained from ED-PIC simulation. |
| electrons.xyz | O | The instantaneous space charge distribution in current ED-PIC step. |

| File | I/O | Description |
|---|---|---|
| emission.dat | O | Saved electron field emission data, including total emission current ($I_t$), mean emission current, maximum local emission current density ($J_{max}$), and maximum local electric field value ($E_{max}$) of simulation steps. |
| exchange-correlation.movie | O | Saved exchange-correlation potential of previous steps. |
| exchange-correlation.xyz | O | Instantaneous exchange-correlation potential of the current step. |
| fields.movie | O | Saved electric field distribution on the tip of previous steps. |
| fields.xyz | O | Instantaneous electric field distribution on the tip in the current step. |
| surface_fields.movie | O | Saved electric field distribution on the tip surface of previous steps. |
| surface_fields.xyz | O | Instantaneous electric field distribution in the current step. |
| FJI.dat | O | Saved maximum local electric field, local emitting current density, and total emitting current for each simulation step. |
| forces.movie | O | Saved electric forces of previous steps for surface atoms. |
| forces.xyz | O | Instantaneous surface electric forces of current step for surface atoms. |
| surface_temperature.movie | O | Saved electron temperature distribution on tip surface for previous steps. |
| surface_temperature.xyz | O | Instantaneous electron temperature distribution on the tip surface. |
| temperature_phonon.movie | O | Electron and phonon temperature distributions of the tip during the evolution. |
| temperature_phonon.xyz | O | Instantaneous electron and phonon temperature distributions of the tip. |
| temperature.dat | O | The saved mean electron and phonon temperature data for all steps. |

**Table C2.** List of keywords in file *edhc.in* for configuring ED+HC simulations. The type, default value, unit, and description for each keyword are provided. (Notations for types: R for real type, I for integer type, S for string, and B for Boolean type.)

| General Parameters | | | | |
|---|---|---|---|---|
| Keyword | Type | Default | Unit | Description |
| infile | S | in/edhc.in.xyz | — | Mesh coordinates of emitter. |
| timestep | R | 4.0 | fs | Time step (should equal to $\Delta t_{MD}$ in ED-MD simulation). |
| timelimit | R | 200.0 | ps | Total simulation time. |
| movie_timestep | I | 1000 | fs | Time interval to write the movie files for atomic coordinates; when the value is set to 0, the file is not written. |
| radius | R | 70.0 | Å | Radius of the coarsening nanotip at the junction between tip and substrate, the value must be larger than the tip radius at the bottom. |
| coarse_theta | R | 10.0 | ° | Apex opening angle of the coarsening nanotip. |
| coarse_factor | R | 0.5 12 2 | — | Coarsening factors (bulk slab, non-apex region, tip apex). For generating dense coarse-graining mesh, the small value should be used for that particular region. |
| Mesh smoothing parameters | | | | |

| Keyword | Type | Default | Unit | Description |
|---|---|---|---|---|
| nnn | I | 12 | — | Number of nearest neighbors for cluster extraction. |
| coord_cutoff | R | 3.1 | Å | Cut-off radius for atomic coordination number analysis. |
| cluster_cutoff | R | 4.2 | Å | Cut-off radius for extracting the vaporized atomic clusters. |
| mesh_quality | R | 1.8 | — | Mesh quality created by Delaunay tetrahedrization; Smaller nr gives more symmetric elements. |
| coplanarity | R | 1.0 | — | Parameter defining the max flatness of tetrahedron. |
| smooth_steps | I | 3 | — | Number of surface mesh smoothing iterations; Setting the value to 0 for avoiding the smoothing step. |
| surface_smooth_factor | R | 1.1 | — | Surface smoothing factor; vary large number may lead to unphysical roughness in the surface mesh. |
| Parameters for coarse-grained domain of tip and electrostatic calculation ||||||
| Keyword | Type | Default | Unit | Description |
| extended_atoms | S | in/extension.xyz | — | Input file with coordinates of surface particles in the coarse-grained extension of nanotip. |
| Z_min | R | -500 | Å | Height of coarse-grained tip section ($h_b$), A negative real number with the origin located at the junction of atomistic model and coarse-grained model. |
| box_width | R | 10.0 | — | Width of simulation box, the multiplication factor with respect to the tip height, i.e., $L_x = L_y = 10h_t$. |
| box_height | R | 10.0 | — | Height of simulation box, the multiplication factor with respect to tip height, i.e., $L_z = 10h_t$. |
| bulk_height | R | 20.0 | — | Thickness of bulk reservoir, the multiplication factor with respect to the lattice constant, i.e., $h = 20a$. |
| work_function | R | 4.59 | eV | Work function of the material. |
| emitter_qe | B | True | — | Quantum effects of space charges are considered when it is set to 'True', otherwise use 'False' for classic field emission model. |
| func_X_id | I | 1 | — | Exchange functional ID. |
| func_C_id | I | 9 | — | Correlation functional ID. |
| temperature_mode | S | Double | — | Options: 'single' (lattice heat conduction) or 'double' (two-temperature model). |
| Heating conduction simulation parameters ||||||
| Keyword | Type | Default | Unit | Description |
| t_ambient | R | 300.0 | K | Temperature of the bulk reservoir of the simulation cell, $T_0$. |
| lorentz | R | $2.0 \times 10^{-8}$ | W·Ω·K$^{-2}$ | Lorentz constant ($L$). |
| heat_cp | R | $3.5 \times 10^{-24}$ | J·K$^{-1}$·Å$^{-3}$ | Volumetric heat capacity of phonons, $C_P$. |
| couplingG | R | $2.0 \times 10^{17}$ | W·K$^{-1}$·m$^{-3}$ | Electron-phonon energy exchange rate, $G_{ep}$. |
| electron_gamma | R | 184.75 | J·K$^{-2}$·m$^{-3}$ | Volumetric heat capacity of electrons, $C_e$. |

| Keyword | Type | Default | Unit | Description |
|---|---|---|---|---|
| Ncell | I | 3 3 100 | — | The rectangular mesh grid for computing temperature profiles from heat balance equations $N_x$, $N_y$, $N_z$. |
| heat_dt | R | 40.0 | fs | Heat calculation time step. |
| ED-PIC simulation with space charge quantum effects | | | | |
| Keyword | Type | Default | Unit | Description |
| elfield_mode | S | DC | — | Type of the applied electric field: DC, AC, pulse, etc. |
| elfield | R | -0.032 | V·Å$^{-1}$ | Value of applied electric field. |
| elfield_rate | R | 10.0 | — | The pulse rate in V·Å$^{-1}$·ns$^{-1}$, AC frequency in GHz, or time constant tau in ns, depending on the mode. |
| elfield_limit | R | -0.3 | V·Å$^{-1}$ | The maximum value for pulses, or the amplitude for AC and exponential fields. |
| Vappl | R | 30 | V | Anode voltage when Dirichlet boundary conditions is used. |
| anode_bc | S | neumann | — | Boundary condition type at anode (Dirichlet or Neumann). Dirichlet condition is current in experimenting and only applies to nanogap simulation. |
| gap_distance | R | 10.0 | nm | Nano-gap spacing must be specified for Dirichlet boundary conditions, i.e., parameter $D$. For Neumann boundary condition when applying space charge quantum effects corrected field emission model, this is an auxiliary parameter, and using the default value $D = 10$ nm is sufficient in most calculations. |
| n_lines | I | 100 | — | Number of discrete points for solving 1D Poisson-Schrödinger equation, $N_{1D}$. |
| alpha | R | 2.0 | — | Normalized field emission cutoff distance in the 1D model, using default value $\alpha = 2.0$ is sufficient in calculations. |
| PIC parameters | | | | |
| Keyword | Type | Default | Unit | Description |
| pic_dtmax | R | 0.5 | fs | Time interval between particle collisions in PIC. |
| electron_weight | R | 0.01 | — | The electron super-particle weight, defined as the number of super-particle. |

**Appendix D. Input and output files for ED-MD-PIC simulation**

For performing a standard ED-MD-PIC simulation, the required input and output files are a combination of those detailed in Appendix B and C. During the simulation, the atomic coordinates of the molecular dynamics simulation are automatically converted into the finite element mesh point file *edhc.in.xyz*. Therefore, users have no need to provide this file separately. The parameter of *ensemble* in *md.in* should be set to *nonEquil* for non-equilibrium molecular dynamics of ED-MD-PIC simulations. Additionally, due to the consecutive movements of atoms in nanotip during MD simulation, it is computationally expensive to constantly update the finite element mesh at every step. Therefore, we have implemented a strategy where we track the root-mean-square distance (RMSD) that the atoms have moved since the last full finite-element calculation [21]. When it exceeds a predefined threshold value, the finite element mesh is regenerated according to the updated atomistic model of nanotip. Otherwise, if RMSD is below the threshold value, the existing mesh is used for electric field calculations. This approach significantly improves computational efficiency without losing accuracy.

$$\text{RMSD} = \sqrt{\frac{1}{N}\sum_{i \in N}\left|\vec{r}_i - \vec{r}_i^{\text{ref}}\right|^2} \tag{D1}$$

Here, $N$ is the atoms number, and the accumulation term is the square of the distance between the current atomic coordinates $\vec{r}_i$ and the reference atomic coordinates $\vec{r}_i^{\text{ref}}$ that are used to construct the finite element mesh in the previous iteration. In that case, the user needs to set the threshold value for the RMSD in the file *fecocs.in*. The value for the RMSD is determined empirically. It determines the interval at which the mesh rebuilding is triggered based on the accumulated displacement of the atoms.

    distance_tol = 0.45        # max RMS distance atoms are allowed to move between runs before the solution is recalculated; 0 enforces the mesh generation for every time-step

During the execution of the ED-MD-PIC model, dynamic outputs of all modules' results are stored in the *out* folder, referring to Appendix B and C for more details.